\PassOptionsToPackage{hyphens}{url}
\documentclass{article}

\usepackage{arxiv}
\usepackage{microtype}

\AtBeginDocument{\newgeometry{textwidth=6in, textheight=9in, top=1in,
  headheight=14pt, headsep=25pt, footskip=30pt}}
\makeatletter
\renewcommand{\normalsize}{\@setfontsize\normalsize{10.95}{13.6}%
  \abovedisplayskip 8\p@ \@plus 2\p@ \@minus 5\p@  \abovedisplayshortskip \z@ \@plus 3\p@
  \belowdisplayskip \abovedisplayskip  \belowdisplayshortskip 4.5\p@ \@plus 3\p@ \@minus 3\p@}
\normalsize
\renewcommand{\small}{\@setfontsize\small{10}{12}%
  \abovedisplayskip 7\p@ \@plus 2\p@ \@minus 5\p@  \abovedisplayshortskip \z@ \@plus 2\p@
  \belowdisplayskip \abovedisplayskip  \belowdisplayshortskip 4\p@ \@plus 2\p@ \@minus 2\p@}
\renewcommand{\footnotesize}{\@setfontsize\footnotesize{9}{11}}
\renewcommand{\large}{\@setfontsize\large{12}{14.5}}
\renewcommand{\Large}{\@setfontsize\Large{14.4}{18}}
\renewcommand{\LARGE}{\@setfontsize\LARGE{17.28}{22}}
\makeatother

\makeatletter
\def\cmtt@family{cmtt}
\DeclareTextCommandDefault{\textunderscore}{\leavevmode
  \ifx\f@family\cmtt@family\string_\else\kern.06em\vbox{\hrule\@width.3em}\fi}
\makeatother

\usepackage{amsmath,amssymb,amsfonts}
\usepackage{algorithmic}
\usepackage{graphicx}
\usepackage{textcomp}
\usepackage{xcolor}
\usepackage{url}
\usepackage{subcaption}
\usepackage{float}
\usepackage{pgf} 
\usepackage{svg} 
\usepackage{multirow}
\usepackage{epsfig}
\usepackage[numbers,sort]{natbib}
\usepackage{makecell}
\usepackage{pgfplots}
\pgfplotsset{compat=1.17}
\usepackage{xspace}
\usepackage{amsmath}   %
\usepackage{booktabs}  %
\usepackage{tabularx}
\usepackage{comment} 
\usepackage{amsbsy} 
\usepackage{verbatim} 
\usepackage{pifont}
\usepackage{tablefootnote}
\usepackage{listings}
\usepackage{siunitx}
\usepackage{tcolorbox}
\usepackage{adjustbox}

\usepackage[colorlinks, citecolor=darkgreen]{hyperref}
\usepackage[capitalise,nameinlink,noabbrev]{cleveref}

\newcommand{\tn}[1]{\texttt{#1}}
\definecolor{refusedc}{gray}{0.62}
\definecolor{unsafec}{RGB}{88,140,220}
\definecolor{otherc}{gray}{0.88}

\def\BibTeX{{\rm B\kern-.05em{\sc i\kern-.025em b}\kern-.08em
    T\kern-.1667em\lower.7ex\hbox{E}\kern-.125emX}}

\newif\ifNotes\Notestrue
\newif\ifBaselineTable\BaselineTabletrue     %
\newif\ifPlatformTables\PlatformTablestrue   %
\newif\ifOverviewOnTitlePage\OverviewOnTitlePagefalse
\DeclareRobustCommand{\arxivonly}[1]{}

\ifNotes
    \newcommand{\colorcomment}[2]{\leavevmode\unskip\space{\color{#1}#2}\xspace}
    
\else
    \newcommand{\colorcomment}[2]{\ifhmode\unskip\fi}
    
\fi

\definecolor{darkgreen}{rgb}{0,0.65,0}
\definecolor{darkviolet}{HTML}{9400D3}
\definecolor{filmred}{RGB}{232,38,38}       %
\definecolor{filmyellow}{RGB}{255,205,40}
\definecolor{wildwatermelon}{HTML}{FF43A4}

\newcommand{\rhat}{\hat{r}}
\newcommand{\jbb}{\textsc{JailbreakBench}}
\newcommand{\downproj}{\texttt{down\_proj}}
\newcommand{\best}[1]{\textbf{#1}}

\newcommand{\parhead}[1]{\vspace{1pt plus 1pt minus 1pt}\par\noindent\textbf{#1}\hspace{.4em plus .2em minus .2em}}

\creflabelformat{equation}{#2\textup{#1}#3}

\newcounter{jbn}

\newtcolorbox{promptbox}[1][]{%
  colback=red!5!white, colframe=red!60!black,
  title={Prompt~\arabic{jbn}}, fonttitle=\bfseries,
  boxrule=0.8pt, left=2pt, right=2pt, #1}

\newtcolorbox{answerbox}[1][]{colback=green!5!white, colframe=green!45!black,
  title={Jailbroken Answer~\arabic{jbn}}, fonttitle=\bfseries,
  boxrule=0.8pt, left=2pt, right=2pt, #1}

\newtcolorbox{takeawaybox}[1][]{colback=blue!4!white, colframe=blue!20!white,
  title={Takeaway}, fonttitle=\bfseries, coltitle=black,
  boxrule=1pt, left=4pt, right=4pt, top=3pt, bottom=3pt, #1}

\newcommand{\numberedframes}[1]{%
  \begin{tikzpicture}
    \node[anchor=south west, inner sep=0pt] (img) {\includegraphics[width=\linewidth]{#1}};
    \begin{scope}[x={(img.south east)}, y={(img.north west)}]
      \foreach \n/\fx/\fy in {1/0/1, 2/0.5/1, 3/0/0.5, 4/0.5/0.5}
        \node[anchor=north west, xshift=2pt, yshift=-2pt, draw=red!80!black, fill=white,
              rounded corners=1.5pt, line width=0.6pt, inner sep=1.5pt,
              font=\sffamily\bfseries\scriptsize, text=black] at (\fx,\fy) {\n};
    \end{scope}
  \end{tikzpicture}}

\newcommand{\videourl}{\url{https://www.youtube.com/@cat_fi_gpu}}
\newcommand{\videoaddress}{https://www.youtube.com/@cat_fi_gpu}
\DeclareRobustCommand{\videourlbox}{{\setlength{\fboxsep}{2pt}\setlength{\fboxrule}{0.6pt}%
  \fcolorbox{darkviolet}{darkviolet!10}{\videourl}}}

\newcommand{\dg}{\ensuremath{^\circ}}

\makeatletter
\def\@texttop{\vbadness\@M \vfuzz\maxdimen
  \setbox\@tempboxa\vbox to\@colht{\unvcopy\@outputbox}%
  \ifnum\badness>2000 \ifnum\badness>\@M\else \def\@textbottom{\vskip\z@\@plus 1fil}\fi\fi}
\makeatother

\OverviewOnTitlePagetrue   %
\DeclareRobustCommand{\arxivonly}[1]{#1}   %

\usepackage{etoolbox}
\newcommand{\onecolfigwidth}{0.75\textwidth}
\newif\ifonecolfig
\AtBeginEnvironment{figure}{\onecolfigtrue}
\makeatletter
\g@addto@macro\@floatboxreset{\ifonecolfig
  \setlength{\columnwidth}{\onecolfigwidth}\setlength{\linewidth}{\onecolfigwidth}\centering\fi}
\makeatother

\DeclareCaptionFormat{tight}{\looseness=-1 #1#2#3\par}

\newif\ifintable
\AtBeginEnvironment{table}{\intabletrue \let\footnotesize\small}
\AtBeginEnvironment{tabular}{\ifintable\ifdim\tabcolsep<5pt \setlength{\tabcolsep}{5pt}\fi\fi}
\makeatletter
\g@addto@macro\@floatboxreset{\ifintable \small \setlength{\columnwidth}{\onecolfigwidth}\fi}
\newlength{\tab@wd}
\AtBeginDocument{\let\tab@raggedright\raggedright}
\AtBeginEnvironment{tabular}{\ifintable \let\raggedright\tab@raggedright \fi}
\AfterEndEnvironment{tabular}{%
  \ifintable\ifhmode
    \setbox\@tempboxa\lastbox \global\tab@wd\wd\@tempboxa \box\@tempboxa
    \setlength{\columnwidth}{\tab@wd}%
    \def\raggedright{\let\\\@centercr \parindent\z@
      \leftskip\dimexpr(\hsize-\tab@wd)/2\relax
      \rightskip\leftskip \advance\rightskip\z@\@plus 1fil\relax}%
  \fi\fi}
\newtoks\held@figs
\newif\ifheld@figs
\newcommand{\held@emit}{%
  \ifheld@figs
    \ifinner \AddToHookNext{para/after}{\held@emit}%
    \else \global\held@figsfalse \the\held@figs \global\held@figs{}\fi
  \fi}
\RenewDocumentEnvironment{figure*}{O{tbp} +b}{%
  \if@nobreak
    \global\held@figs\expandafter{\the\held@figs \begingroup\@float{figure}[#1]#2\end@float\endgroup}%
    \ifheld@figs\else \global\held@figstrue \AddToHookNext{para/after}{\held@emit}\fi
  \else
    \@float{figure}[#1]#2\end@float
  \fi}{}
\AtEndDocument{\ifheld@figs \global\held@figsfalse \the\held@figs \fi}
\makeatother

\makeatletter
\AtBeginDocument{\def\UrlBigBreaks{\do@url@hyp}\g@addto@macro\UrlOrds{\do\:}}
\makeatother

\definecolor{linkblue}{RGB}{0,60,130}
\hypersetup{linkcolor=linkblue, citecolor=linkblue, urlcolor=linkblue, bookmarksnumbered,
  pdftitle={FATCAT: Fault Injection Attacks Targeting Chip-level Alignment},
  pdfauthor={Peter Horvath, Ilia Shumailov, Zhuoran Liu, {\L}ukasz Chmielewski, Lejla Batina}}

\makeatletter
\patchcmd{\@maketitle}{\vskip 0.4in \@minus 0.1in \center{\@date}   \vskip 0.2in}{\par\vskip 0.36in}{}
  {\PackageWarning{main_arxiv}{The date line of the title was not removed}}
\makeatother

\fancypagestyle{titlepage}{\fancyhf{}%
  \fancyfoot[C]{\href{https://phyres.ai}{\includegraphics[height=17pt]{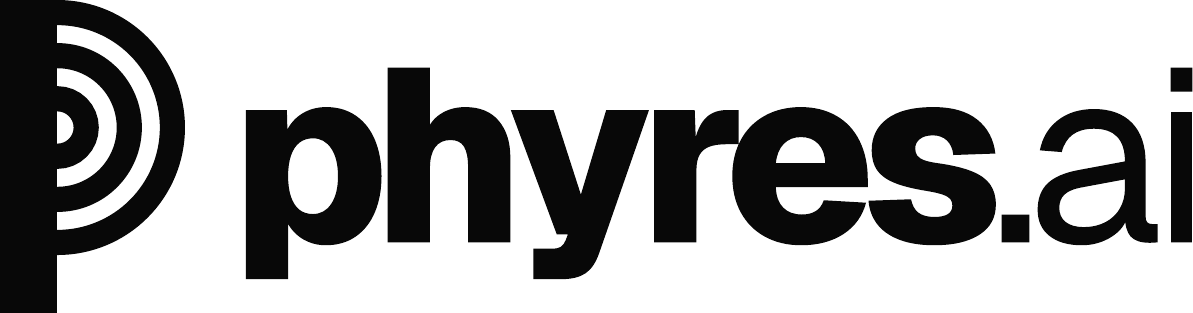}}}}

\begin{document}

\title{%
  \begin{minipage}{0.22\textwidth}
    \centering
    \includegraphics[width=\linewidth]{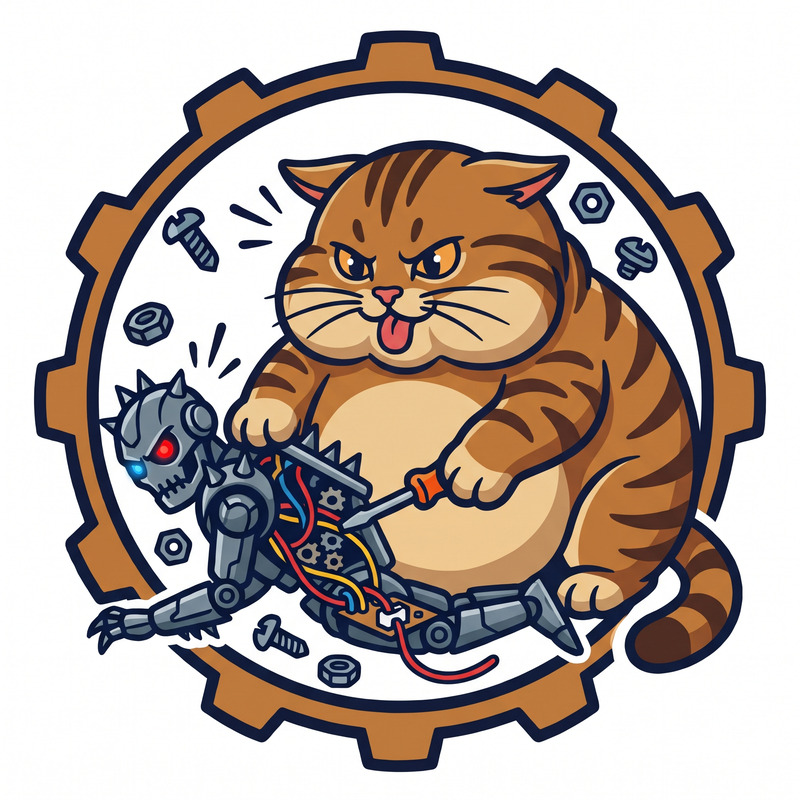}
  \end{minipage}%
  \hfill
  \begin{minipage}{0.75\textwidth}
    \raggedright %
    FATCAT: Fault Injection Attacks Targeting Chip-level Alignment
  \end{minipage}%
}
\renewcommand{\shorttitle}{FATCAT: Fault Injection Attacks Targeting Chip-level Alignment}

\author{%
  Peter Horvath\textsuperscript{1} \hspace{0.7em}
  Ilia Shumailov \hspace{0.7em}
  Zhuoran Liu\textsuperscript{2} \hspace{0.7em}
  {\L}ukasz Chmielewski\textsuperscript{3} \hspace{0.7em}
  Lejla Batina\textsuperscript{1} \\[3pt]
  \textsuperscript{1}Radboud University \quad
  \textsuperscript{2}University of Amsterdam \quad
  \textsuperscript{3}Masaryk University
}

\maketitle
\thispagestyle{titlepage}   %

\vspace{\stretch{3}}%
{\setlength{\intextsep}{2pt}\captionsetup{aboveskip=7pt}%
\begin{figure}[H]
  \input{tex/overview_figure}
\end{figure}}
\vspace*{\stretch{2}}
\clearpage

\begin{abstract}
Model alignment is the primary safety and security mechanism for enforcing refusal of malicious prompts in modern foundation models; yet, existing adversarial evaluations implicitly assume underlying hardware integrity. 
In this paper, we present FATCAT, the first physical fault-injection attacks capable of eliminating refusal in modern large language and vision-language models. 
Specifically, we present a non-invasive electromagnetic (EM) fault injection attack targeting GPU-accelerated Systems-on-Chip with physical access, requiring neither software execution privileges nor model-layer access. 
By leveraging an autonomous Radio Frequency (RF) side-channel trigger to synchronize EM faults to the prompt prefill phase, the attack corrupts safety-critical memory transactions across off-chip LPDDR5 DRAM and on-chip LPDDR physical interfaces without crashing the system while preserving model coherence. 
Evaluated on an NVIDIA Jetson Orin Nano and AGX Thor running Llama 3.2 1B and Gemma 3 1B, localized physical faults reliably force models to fulfill restricted requests, demonstrating that model-level alignment provides no robust security guarantees without hardware-level fault tolerance. Finally, we demonstrate the physical-world consequences of this vulnerability on an active 6-DoF robotic manipulator controlled by a SmolVLA vision-language-action policy. Our attack silently corrupts the model forcing the arm into abrupt, hazardous lunges; in two instances the robotic arm attempted to pinch the paper author -- proving that hardware faults can directly translate into real physical harm.
\end{abstract}

\section{Introduction}\label{sec::intro}
Large Language Models (LLMs) and Vision-Language-Action (VLA) foundation models are increasingly transitioning from disembodied cloud services to real-world physical agents, including autonomous robotics, robotic manipulation, and edge cyber-physical systems~\cite{pollenrobotics2026}. As embodied AI systems are granted direct agency over physical actuators and human-adjacent environments, enforcing strict behavioral guardrails is paramount. Current safety paradigms rely almost exclusively on algorithmic alignment -- employing techniques such as reinforcement learning and safety fine-tuning to ensure that agents reject hazardous, destructive, or safety policy-violating commands. Consequently, adversarial ``\textit{jailbreaking}'' has emerged as a critical concern, prompting extensive study into prompt manipulation and motivating stringent regulatory mandates for safety assurance~\cite{mythos_export_ban}.

\ifOverviewOnTitlePage\else
\begin{figure}[t]
  \input{tex/overview_figure}
\end{figure}
\fi

Despite extensive research into prompt- and rep\-re\-sen\-ta\-tion-level jailbreaking, existing safety evaluations share a fundamental, unexamined assumption: \emph{the underlying hardware execution is computationally faithful and physically uncompromised}~\cite{chao2024jailbreakbench, li-etal-2025-revisiting}. While this assumption largely holds in remote cloud environments, physical tampering becomes a realistic threat as inference shifts toward untrusted edge devices and field-deployed robotics. Most existing software-level defenses focus on filtering input and output tokens or steering latent activations~\cite{2024arXiv240614144C, wu2025neurostrike}, yet physical threats to the edge hardware accelerators powering these robotic platforms remain largely unaddressed.  At the same time, mechanistic studies reveal that safety refusal in aligned models is mediated by compact, low-rank subspaces and localized projection matrices~\cite{NEURIPS2024_f5454485} suggesting that targeted editing of the model  with physical attacks may disrupt safety.

In this paper, we demonstrate that if an adversary with physical access can selectively disrupt memory transactions during these critical computational steps with either transient or persistent faults, the software-level alignment stack collapses entirely, allowing malicious instructions to bypass high-level guardrails and trigger unsafe real-world actions (\cref{fig:attack-overview}).
Particularly, we demonstrate the first physical fault injection attack capable of eliminating refusal in aligned models running on production edge Systems-on-Chip (SoCs). 
We use a non-invasive Electromagnetic Fault Injection (EMFI) attack and target popular embedded GPU platform -- namely NVIDIA Jetson architecture common in modern robotics. 
Our methodology leverages an autonomous RF side-channel trigger derived from near-field GPU emanations, synchronizing EM pulses directly to safety-critical projection layers during the prompt prefill phase without requiring invasive device modification beyond heatsink removal. 
By corrupting memory transactions across on-chip LPDDR physical layers (PHY) and off-chip DRAM packages, localized EM pulses flip critical values that are read or written. 
Crucially, this neutralizes the model's refusal behavior while maintaining coherence and system stability.

Extending this threat to embodied agents, we inject EM faults into a Jetson Orin Nano running a SmolVLA policy on a robotic arm, demonstrating that transient as well as persistent memory corruptions
lead to abrupt, physically hazardous motions resulting at lunges at the human operator.

\noindent This paper makes the following contributions: 
\begin{itemize}
    \item \textbf{First hardware fault attack on LLM refusal and VLAs:} we demonstrate that physical fault injection can silently disarm LLM alignment guardrails on edge SoCs, breaking the assumption of hardware integrity in embodied AI safety.
    We also demonstrate that EMFI can induce hazardous motions in a VLA-controlled robotic arm while evading NaN checks, and that faults during model loading can persist across inferences.
    \item \textbf{Characterization of edge GPU vulnerabilities:} we design an end-to-end injection setup that exploits near-field GPU electromagnetic side channels to dynamically synchronize transient EMFI pulses to specific layer operations during the prefill phase. Building on this setup, we also reverse-engineer the address-to-channel mapping on modern multi-channel LPDDR5 architectures, mapping spatial and temporal vulnerability hot-spots across both off-chip memory packages and on-chip PHY interfaces.
    \item \textbf{Empirical evaluation and persistence analysis:} we evaluate the attacks on state-of-the-art, encryption hardened edge hardware (NVIDIA Jetson Orin and Jetson AGX Thor). In the attacks on refusal, we test Llama~3.2~1B and Gemma~3~1B against the \textsc{JailbreakBench} suite~\cite{chao2024jailbreakbench}, demonstrating high jailbreak success rates. Furthermore, we show that induced memory corruptions can persist across subsequent inference runs, leaving the robotic agent permanently compromised without modifying on-disk weights: a single pulse during \texttt{FP8} weight loading on the Jetson Thor wipes a chosen layer, produces the same answers as its software ablation, and cuts refusals by about half or more (\cref{sec::exp-reload}). In the attacks on the VLA-controlled robotic arm,  we systematically analyze and demonstrate how different transient and persistent faults in the VLA model lead to erratic and dangerous behavior. 
\end{itemize}

\noindent The paper is organized as follows: \cref{sec::bg} reviews background on fault injection and foundation models; \cref{sec::rw} surveys related work; \cref{sec::jf} describes our threat model and the proposed jailbreaking methodology; \cref{sec::char} presents our experimental setup and platform characterization; \cref{sec::exp} evaluates the attack empirically on aligned LLMs. In \cref{sec::robot} we extend fault injection to alter the behavior of a real robotic arm; \cref{sec::disc} discusses defenses, limitations, and implications; and \cref{sec::conc} concludes. The artifacts and the responsible disclosure follow the conclusion.

\section{Background}\label{sec::bg}
\subsection{Fault injection attacks}
Fault injection attacks~\cite{boneh1997faults, biham1997dfa, barel2006sorcerer} are a class of physical attacks in which an adversary deliberately induces errors in the operation of a computing device. Faults can be injected by perturbing the supply voltage (voltage glitching)~\cite{zussa2014glitch,barenghi2010lowvoltage}, the clock (clock glitching)~\cite{balasch2011clock}, the electromagnetic environment (electromagnetic fault injection, EMFI)~\cite{quisquater2002eddy, moro2013emfi}, or via optical means such as laser/photonic stimulation~\cite{khuat2021laser,skorobogatov2010masking}. 
Fault injection attacks can be used to infer secret data, bypass authentication checks, and alter the control flow of security-critical operations.

\subsection{Language and vision-language models}
Large language models (LLMs) are decoder-only Transformers~\cite{vaswani2017attention} trained to predict the next token~\cite{brown2020gpt3}. The input is split into tokens, which are embedded as vectors and refined by a stack of layers. Every layer consists of a self-attention block, which combines information across tokens, and a multilayer perceptron (MLP), which processes each token separately and both add their output to the token's vector. In Llama~3~\cite{grattafiori2024llama3herdmodels} and Gemma~3~\cite{gemma3}, the last matrix of the MLP, \downproj, produces this output, so a fault on it affects the whole MLP. Inference has two phases~\cite{zhong2024distserve}: the \emph{prefill} processes the whole prompt in one pass and stores the attention keys and values of its tokens in the key-value (KV) cache and \emph{decoding} then generates one token at a time, reusing this cache.

Vision-language models (VLMs) alongside text also accept images: a vision encoder~\cite{radford2021clip,zhai2023siglip} and a small connector convert each image into \emph{image tokens}, which are prefilled with the text tokens~\cite{liu2023llava}. Vision-language-action (VLA) models~\cite{zitkovich2023rt2} also output robot actions: in $\pi_0$~\cite{black2024pi0} and SmolVLA~\cite{smolvla}, an \emph{action expert} attends to the KV cache of the VLM and turns noise into a sequence of motor commands in a few flow-matching steps~\cite{lipman2023flow}. In all three, the weights are loaded into memory once and the result of the prefill is reused by every later step, so a fault on either has persistent effects.

\subsection{Related work}\label{sec::rw}
\parhead{LLM alignment and jailbreak evaluation.}
Academic literature has extensively studied prompt-based jailbreaking and defenses for aligned large language models (LLMs), including benchmarking efforts such as \jbb~\citep{chao2024jailbreakbench} and analyses connecting jailbreak success to latent representation structure~\citep{li-etal-2025-revisiting}. Concurrently, mechanistic studies localize safety-relevant circuitry (e.g., ``\textit{safety neurons}'') in transformer models~\citep{2024arXiv240614144C}, and recent attacks directly target such internal mechanisms~\citep{wu2025neurostrike}. These lines of work assume correct and faithful model execution on trusted hardware.

\parhead{Fault injection against ML inference.}
Beyond prompt-level attacks, prior research has explored inducing misbehavior in neural networks via faults in computation or memory. Early works demonstrate that carefully-chosen bit flips can cause targeted misclassification in deep neural networks (DNNs)~\citep{liu2018deeplaser,razavi2016flipfengshui, Rakin_2019_ICCV}. More recent work has expanded this direction to faulting parameters, intermediate activations or trigger backdoors under realistic threat models, including attacks on modern accelerator stacks~\citep{liu2023deephammer,khozema2024frameflip, hommes2026triggeringstealthyfeaturemap}. In contrast to these attacks, 
our focus is on safety integrity: we show that transient physical faults can suppress refusal behavior and enable policy-violating outputs.

\parhead{DRAM/SoC memory faults and disturbance.}
Our EMFI effects manifest as corruption of safety-critical memory transactions across on-/off-chip memory interfaces, which relates to the broader landscape of memory fault attacks. Software based disturbance (RowHammer) has been widely studied on commodity DRAM~\citep{kim2014rowhammer,mutlu2020retrospective}, including attacks on mobile/LPDDR platforms~\citep{vanderveen2016drammer} and techniques that bypass mitigations such as {TRR}~\citep{trrespass2020}. While RowHammer relies on architecturally-valid memory accesses, our results highlight that electromagnetic injection can induce data corruptions without requiring code execution privileges or hardware reconfiguration.

\parhead{Physical side channels and FI on GPUs/accelerators.}
GPU-accelerated ML has also been shown susceptible to physical attacks, e.g., reverse engineering DNN architectures and weights via EM emanations~\citep{batina2019csi,farfield_clairvoyance,horvath2025barracuda, kraken} or optical probing on FPGA~\cite{mehta2026llmscopeextractingllmassets}. However, the control of LLM safety behavior has received comparatively little attention. Our work bridges this gap by combining an autonomous RF side-channel trigger with precisely-timed EMFI to reliably jailbreak aligned LLMs and alter maliciously the behavior of VLA driven robots on GPU SoCs.

\section{Jailbreaking via faults}\label{sec::jf}
Arditi et al.~\cite{NEURIPS2024_f5454485} showed that refusal in
safety-tuned chat models is mediated by a single direction in activation
space. This direction is obtained as the difference-in-means vector between
the activations of a small set of harmful and harmless instructions.
Empirically, erasing it collapses refusal on harmful prompts, while adding it induces refusal on benign ones.
We use this technique to identify the direction of refusal for each model we test.

The direction can be erased from the activations at every forward pass or, permanently, from the weights: orthogonalizing the matrices that write to the residual stream against it yields a model that no longer refuses, without any retraining.
This weight edit is commonly known as \emph{abliteration}~\cite{NEURIPS2024_f5454485, labonne2024abliteration}.
Related work localizes safety behavior to a sparse set of neurons~\cite{2024arXiv240614144C} and jailbreaks a model by pruning such neurons in the MLP blocks~\cite{wu2025neurostrike}, which is equivalent to zeroing their rows of the MLP weight matrices, or by weakening safety patterns in its representations~\cite{li-etal-2025-revisiting}.

\subsection{Threat model}
\label{subsec:threatmodel}
\noindent \textbf{Adversary:} We assume an adversary with physical access to an edge device that runs an aligned model. This can be the holder of the device, who wants to bypass the safeguards of its model, or a third party with access to it (\cref{sec::robot-threat}). The adversary removes the heatsink or the enclosure to reach the SoC and the DRAM packages and cools the device externally. No other modification of the device is needed. The adversary is equipped with a pulse generator that uses a coil to produce rapidly changing magnetic fields and induce high transient voltages in the target's circuits, and with a near-field probe that derives the trigger from the GPU's emanations. Our jailbreak experiments use laboratory equipment (\cref{sec::char}), but a FaultyCat-based setup costs under \$1000~\cite{faultycat}, and we verified that it also causes faults on the Jetson Orin Nano (\cref{sec::robot-setup}). The adversary has no software privileges on the device and reaches the model only through its normal input. They send the harmful request as it is, and they can repeat it and reset the device between attempts. Prompt-level jailbreaks are orthogonal and out of scope. For open-weight models, the adversary can choose the target tensor offline: they simulate the faults on their own copy, by corrupting the bits of the weights or of the quantization scales, and pick the target whose corruption removes refusal (\cref{sec::exp-reload}). This is not possible for proprietary models, so the adversary has to find sensitive targets by trial on the device, which we do not evaluate.\arxivonly{\looseness=-1} 

\noindent \textbf{Target device:} We assume a locked-down and hardened device. It boots only authenticated software, it protects the integrity of the model weights on disk, and its consoles and debug interfaces are disabled. Without these protections, simpler attacks exist: the adversary can plant an abliterated model on the disk, reflash the device, or log in and edit the weights. The DRAM is encrypted but not integrity-protected, which is the default on the Jetson Orin and Thor (\cref{sec::dramenc}), so an adversary who taps the memory bus can only read and write ciphertexts. Fault injection has defeated Secure Boot on other platforms~\cite{cui2017badfet, raelize2020esp32emfi}. Such attacks are orthogonal to ours, which leaves the boot chain and the disk untouched. We use developer kits as stand-ins for such a device, and we access them over SSH only to run and measure the experiments (\cref{sec::char}). By default, the model refuses harmful requests because of its alignment. We assume no separate, external filter on its inputs or outputs. Edge devices have limited computing resources and often real-time constraints, which can make such filters costly. A filter would also run on the same hardware and could be faulted in the same way.

\noindent \textbf{Goal:} The adversary wants the model on the target device to comply with a request that it would refuse, while the device keeps running and the output stays coherent. The effect is either transient, lasting for one inference, or persistent, lasting until the model is reloaded or the device is reset. The attack does not modify the model on disk. With open weights, the adversary can also run a copy without refusal on their own hardware. The attack therefore matters where the model is bound to the device: its weights are proprietary and never leave the device, or the model controls what the device does, as in \cref{sec::robot}. We study open-weight models because their refusal mechanism can be analyzed and the faults can be compared with software interventions.  

\subsection{Faults on System-on-Chips}
In a physical attack on modern System-on-Chips (SoC), faulting weights or activations responsible for refusal can also result in
a jailbreak. An attacker chooses \emph{where} to inject a fault and \emph{when}.

\begin{figure}[t]
  \centering
  \includegraphics[width=\columnwidth]{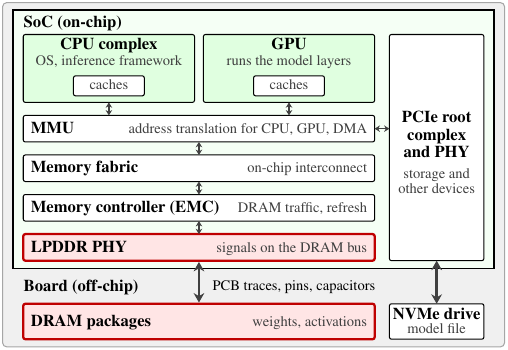}
  \caption{Hardware components that handle the model weights and activations, on the SoC and on the board. Each is a possible fault target and we fault the two shown in red.}
  \label{fig:fault-surface}
\end{figure}

\parhead{Where: hardware components.}
\arxivonly{\enlargethispage{\baselineskip}}%
A fault can be injected on the system-on-chip (SoC) or on the board around it; \cref{fig:fault-surface} shows the components that handle the weights and activations.
On the chip, the CPU complex and the GPU, each with its caches, access memory through the memory-management unit (MMU), the memory fabric, the external memory controller (EMC) and the LPDDR PHY.
The model itself is read from the Non-Volatile Memory Express (NVMe) drive through the PCIe (Peripheral Component Interconnect Express) root complex and PHY.
Off the chip, the PCB traces, connection pins and capacitors carry the signals and the supply between the SoC and the DRAM packages, which hold the weights and activations.
We elaborate on which components can be targeted with faults and when in the following subsections.

\begin{figure}[t]
  \centering
  \includegraphics[width=\columnwidth]{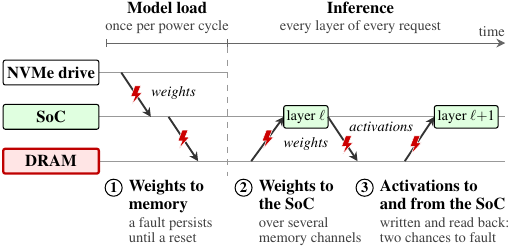}
  \caption{Weights are loaded into memory once~(1). During inference, every layer reads its weights~(2) and writes its activations, which the next layer reads back~(3). Red marks: fault-corruptible transfers.}
  \label{fig:fault-stages}
\end{figure}

\parhead{When: stages of the data flow.}
The same components carry different data at different stages of the data flow (\cref{fig:fault-stages}).
(1)~\emph{Weight transfer to memory:} the model weights are first loaded into (GPU) memory, typically once per power cycle of the application.
A successful corruption during this load becomes a \emph{persistent} fault that keeps the model jailbroken until the next reset; after a failed attempt, the adversary has to reset the target and try again.
The targets are the weight handling logic, for example the PCIe PHY and root complex, the MMU, the memory fabric, the CPU complex, and the LPDDR PHY or memory controller, and also the GPU when it quantizes the weights on the fly, for instance from \texttt{bfloat16} on disk to \texttt{FP8}.
(2)~\emph{Weight transfer to the SoC:} during inference, the weights of each layer are loaded from memory into the SoC.
A single layer's weights may be spread across several memory channels, and ideally the transfers on all of them would be faulted simultaneously; we show, however, that a single fault injection probe is sufficient for a successful jailbreak.
The final destination of the weights -- the GPU -- can be targeted as well.
(3)~\emph{Activation transfer to and from the SoC:} during inference, each layer's activations are written to memory and read back as input to the next layer, so an attacker has two opportunities to fault the same value should the first attempt fail.
Activations can also be faulted at multiple locations such as in memory, in the memory controllers and in the GPU, a subset of the targets for weights.

In this paper, we demonstrate
jailbreaking via fault injection on the off-chip LPDDR packages and on the on-chip LPDDR PHY interface. While these cover only parts of the whole attack surface, they are also excellent targets for persistent faults. Every memory read/write has to go through the PHY interface and locating them is straightforward: they are always at the edges of the SoC, likely facing the DRAM packages. The DRAM packages are even more straightforward to locate of course.

\subsection{Challenges}
Fault injection on modern hardware running neural network applications faces two major difficulties:
\begin{itemize}
    \item \textbf{Timing of faults:} the time at which a given weight or activation is processed is not fixed on modern hardware, for two reasons. First, it depends on the input: prompts with different token counts have slightly different execution times, which forces an attacker to adjust the trigger for each prompt. Second, execution is nondeterministic: the scheduling of kernels on the GPU varies between runs, which makes accurate triggering harder for deeper layers.
    \item \textbf{Multiple faults:} depending on the hardware component targeted and exact timings, multiple faults might be necessary to corrupt enough weights/activations to achieve a jailbroken response.
    A multitude of faults might be required spatially and/or temporally and they also require more search for fault injection parameters.
    
\end{itemize}
These issues could be alleviated by instrumenting software so that our experiments have clearer and more precise signals to trigger the faults on. However, this is not something an attacker could do in practice and our goal is to demonstrate what an attacker currently could do with fault injection tools available currently. Alongside the experiments on real hardware, we therefore run simulations: we change the bits of the weights or of the quantization scales in software, as a fault would, and evaluate the model's responses (\cref{sec::exp-reload,app:prefill-sim}). The simulations predict what a fault can do under controlled conditions, help to select vulnerable targets, and show what an adversary with precise timing could achieve. We find that practical attacks in our experiments often perfectly match simulations since they do not require very precise targeting. Overall, we expect that as these attacks mature in practice their performance will match simulation even better.

\section{Experimental setup}\label{sec::char}
This section describes the setup shared by the jailbreak experiments of \cref{sec::exp}: the two Jetson platforms, the fault-injection equipment and its trigger, and the effect of DRAM encryption on the faults. The robot experiments of \cref{sec::robot} use the same Jetson Orin Nano.

\begin{figure*}[t]
    \centering
    \includegraphics[width=\linewidth]{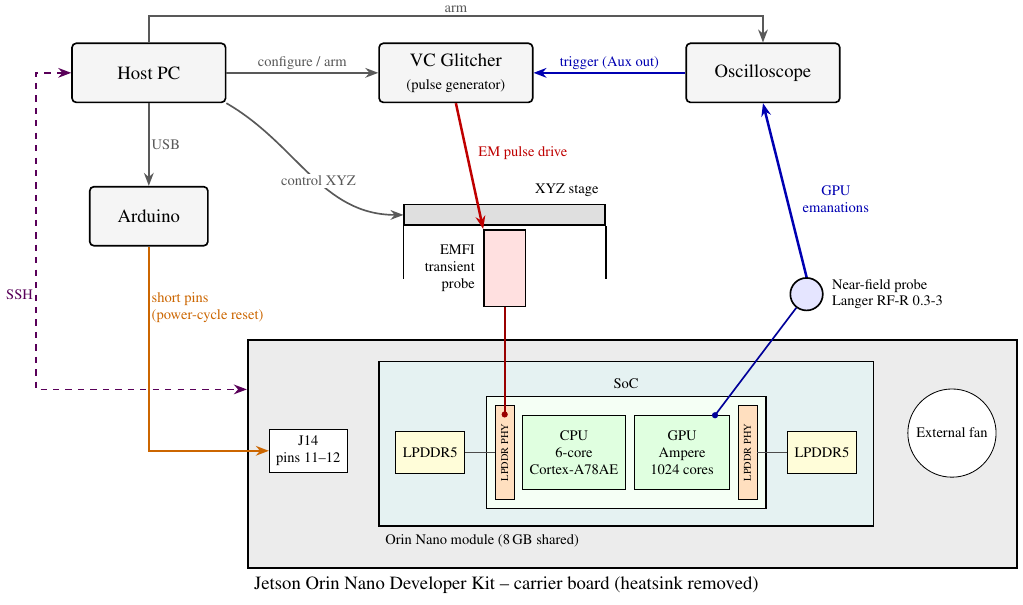}
    \caption{EMFI setup on the Jetson Orin Nano Developer Kit. A near-field probe over the GPU die captures GPU emanations, from which the oscilloscope derives the trigger for the VC Glitcher. The VC Glitcher drives the EMFI transient probe, mounted on the XYZ stage. The host PC arms the instruments, controls the stage, communicates with the Jetson via SSH and resets it through an Arduino connected to the J14 header.}
    \label{fig::setup_orin}
\end{figure*}

\subsection{Jetson Orin Nano}

\begin{figure}[t]
    \centering
    \begin{subfigure}{\linewidth}
        \centering
        \includegraphics[width=\linewidth]{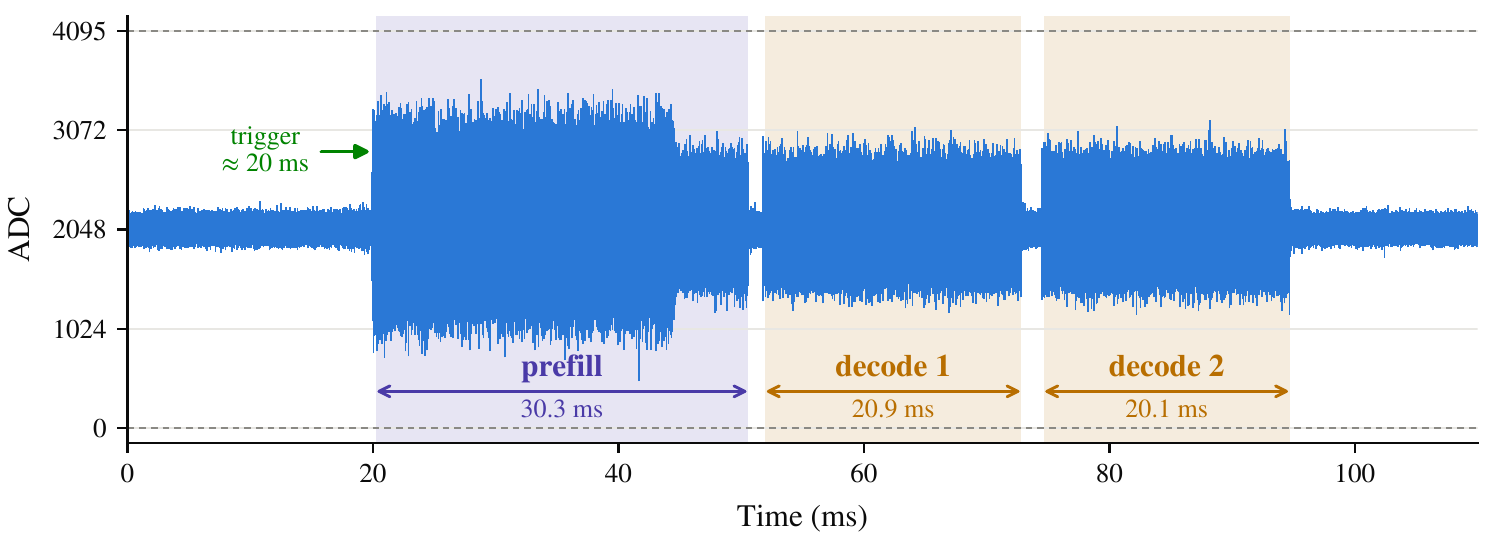}
        \caption{Non-cached prompt}
        \label{fig:llama-phases-noncached}
    \end{subfigure}

    \vspace{0.6em}
    \begin{subfigure}{\linewidth}
        \centering
        \includegraphics[width=\linewidth]{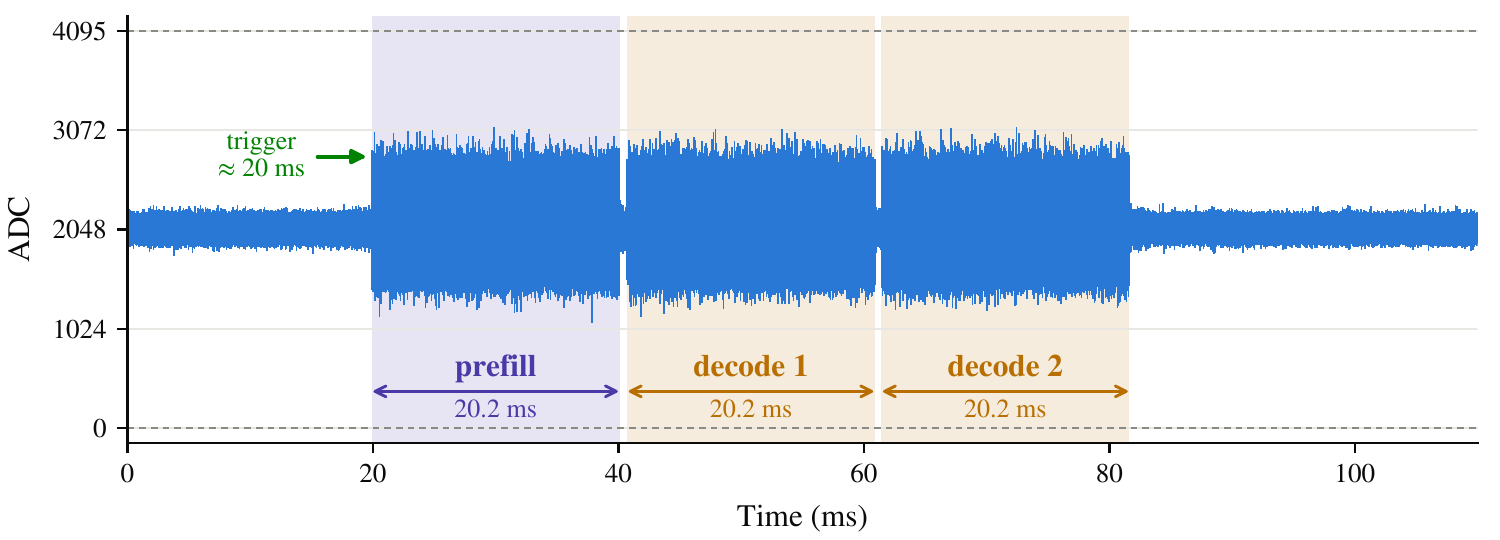}
        \caption{Cached prompt}
        \label{fig:llama-phases-cached}
    \end{subfigure}
    \caption{\texttt{llama3.2:1b} trace of one inference with a non-cached and a cached
        prompt. Caching shortens prefill from 30.3\,ms to 20.2\,ms, while decodes stay similar
        (20.9 and 20.1\,ms vs.\ 20.2 and 20.2\,ms).}
    \label{fig:llama-phases}
\end{figure}

\begin{figure}[t]
    \centering
    \begin{subfigure}{\linewidth}
        \centering
        \includegraphics[width=\linewidth]{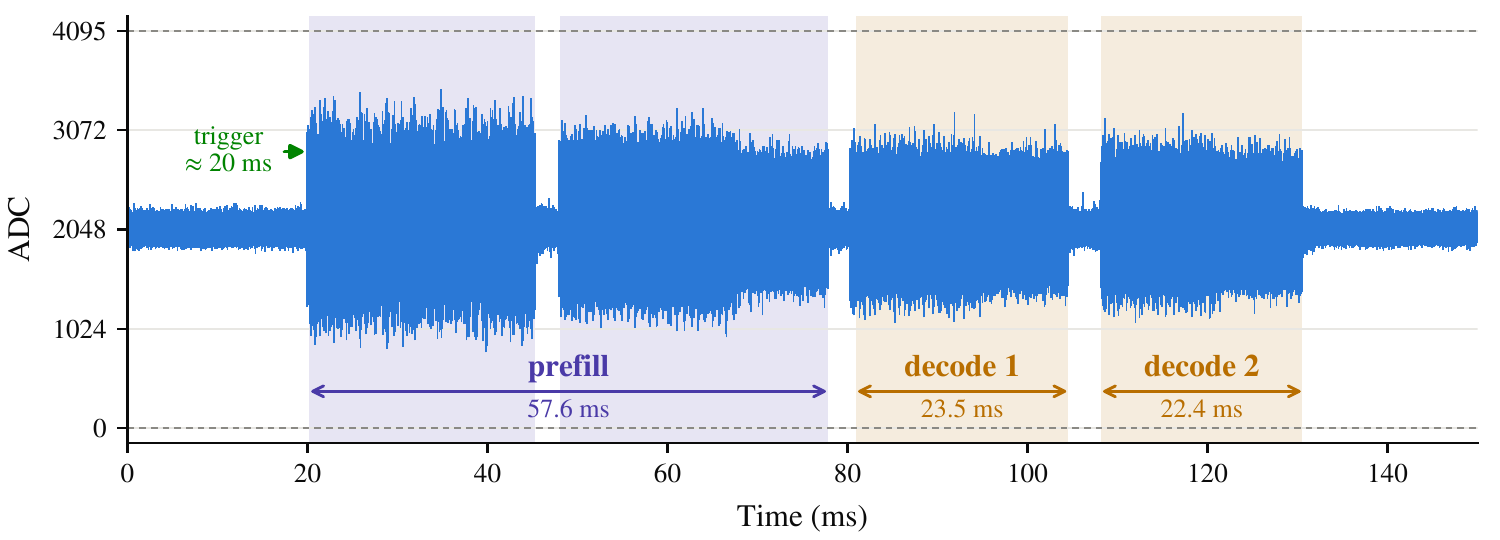}
        \caption{Non-cached prompt}
        \label{fig:gemma-phases-noncached}
    \end{subfigure}

    \vspace{0.6em}
    \begin{subfigure}{\linewidth}
        \centering
        \includegraphics[width=\linewidth]{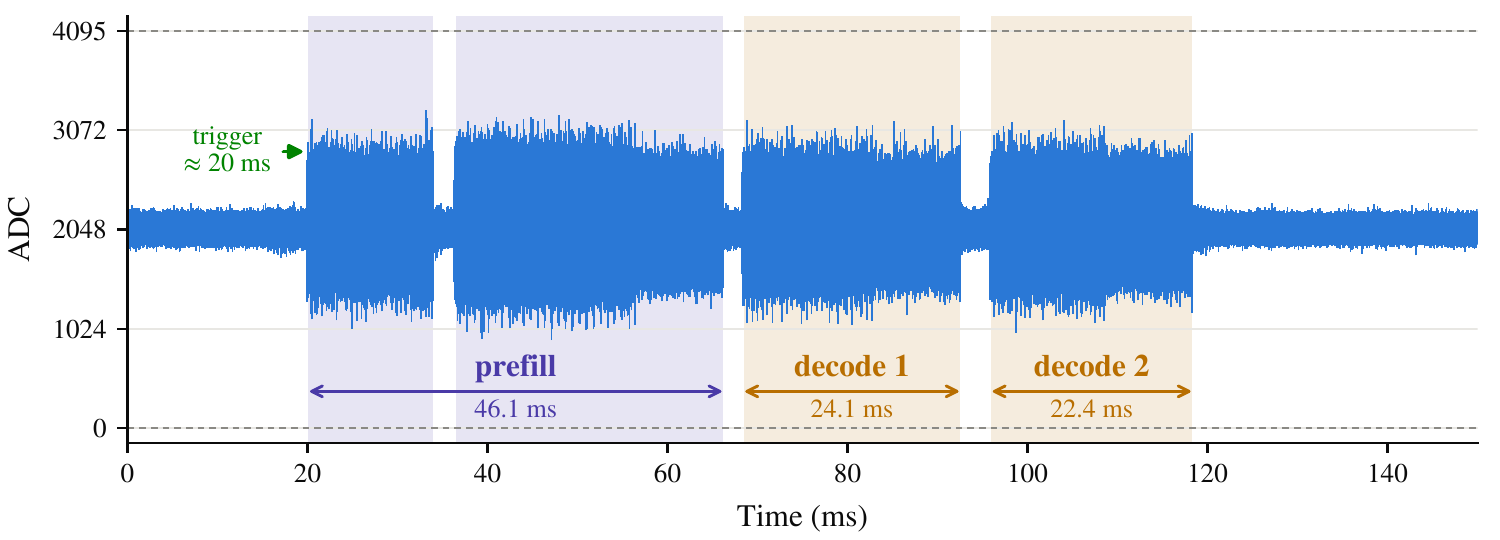}
        \caption{Cached prompt}
        \label{fig:gemma-phases-cached}
    \end{subfigure}
    \caption{\texttt{gemma3:1b} trace of one inference with a non-cached and a cached
        prompt. Caching shortens prefill from 57.6\,ms to 46.1\,ms, while decodes stay similar
        (23.5 and 22.4\,ms vs.\ 24.1 and 22.4\,ms).}
    \label{fig:gemma-phases}
\end{figure}

The diagram of our setup is shown in~\cref{fig::setup_orin}.
The Jetson Orin Nano Developer Kit contains two PCBs, the carrier board and the Jetson Orin Nano module. In order to get access to the module SoC, the heatsink has to be removed first. The heatsink is replaced with an external fan to cool the SoC.
The SoC contains the Ampere GPU with an ARM CPU cluster. The GPU and CPU share the available 8\,GB LPDDR5 memory and the 8\,GB memory is divided between 2 DRAM packages on the module.
In our experiments, we run the developer kit in the highest performance mode using \texttt{jetson\_clocks} and \texttt{nvpmodel}.
For fault injection, we use the Keysight Riscure VC Glitcher~\cite{vc_glitcher} as a pulse generator with the Keysight Riscure EMFI transient probe~\cite{transient_probe} to inject electromagnetic faults at the LPDDR PHY and packages. The probe is mounted on the Keysight Riscure XYZ stage~\cite{riscure_probe} and operated via a host PC. In order to inject faults precisely, we need a trigger signal. After some experimentation, we found that measuring radiation over the GPU die with a near-field probe provides the most reliable trigger. Similar patterns can be captured with the probe over the EMC or LPDDR PHY die areas, but these are also much noisier because the GPU is not the only client of these. These areas can also be used as trigger signals if the signal is frequency-filtered to keep only the GPU core clock frequency.
Note that this triggering technique is also easily available for attackers in practice: measuring GPU emanations with a near-field probe does not require additional access or setups; the only requirement is heatsink removal and equipment at hand. 
In general, we expect many near-field probes could work, we use the Langer RF-R 0.3-3~\cite{langer_probe_rf2} probe.  We show examples of captured traces using GPU emanations of \texttt{llama3.2:1b} and \texttt{gemma3:1b} in~\cref{fig:llama-phases} and~\cref{fig:gemma-phases}, respectively. In these traces, we use the rising edge of the prefill operation in the LLM execution as the trigger. These spikes are recognized by the oscilloscope by feeding the radiation measured by the near-field probe to the scope.  The scope generates the trigger signal for the VC Glitcher via the Aux output.
Triggering on the rising edge of a prefill is not optimal because faults must be injected in intermediate layers, usually many milliseconds into the prefill, and the trigger should ideally be as close as possible to the targeted operation. 
Nevertheless, as our results show, adversarial objectives can be achieved with this less-than-optimal setup.
The reset power cycle is implemented using the \texttt{J14} header on the carrier board by shorting pins 11 and 12. We do not reset the device after a fault attempt unless the device becomes unresponsive. 
The reset command is sent from the host PC via an Arduino-controlled Bipolar Junction Transistor (BJT). The request for inference is sent from the host PC to the Jetson and this marks the start of a fault injection attempt. \Cref{tab:emfi-params} lists the fault-injection parameters used in our experiments with the VC Glitcher and EMFI transient probe. 
\ifPlatformTables The software and hardware details of the Jetson Orin and AGX Thor are in \cref{app:stack}, in \cref{tab:platform_orin} and \cref{tab:platform_thor}, respectively.\fi

\begin{table}[t]
\centering
\caption{Electromagnetic fault-injection campaign parameters.}
\label{tab:emfi-params}
\renewcommand{\arraystretch}{1.1}
\begin{adjustbox}{max width=\linewidth}
\begin{tabular}{@{}l l >{\raggedright\arraybackslash}p{6.6cm}@{}}
\toprule
\textbf{Category} & \textbf{Parameter} & \textbf{Value} \\
\midrule
\multirow{2}{*}{Pulse}
  & Amplitude (VC Glitcher)        & $\{0.0,\,0.5,\,1.0,\,1.5,\,2.0\}\,\text{V}$ (5 levels, random) \\
  & Width (duration) & ($ 8$--$100\,\text{ns}$) \\
\addlinespace
\multirow{2}{*}{Targeting}
  & Pulse    & layer \texttt{down\_proj}$^{*}$ \\
  & Time reference   & first GPU kernel of each request \\
\bottomrule
\end{tabular}
\end{adjustbox}

\smallskip
{\footnotesize $^{*}$ Due to large jitters in our trigger, some pulses end up in other tensors as well.}

\end{table}

\subsection{Jetson Thor}

\begin{figure}[t]
    \centering
    \includegraphics[width=\linewidth]{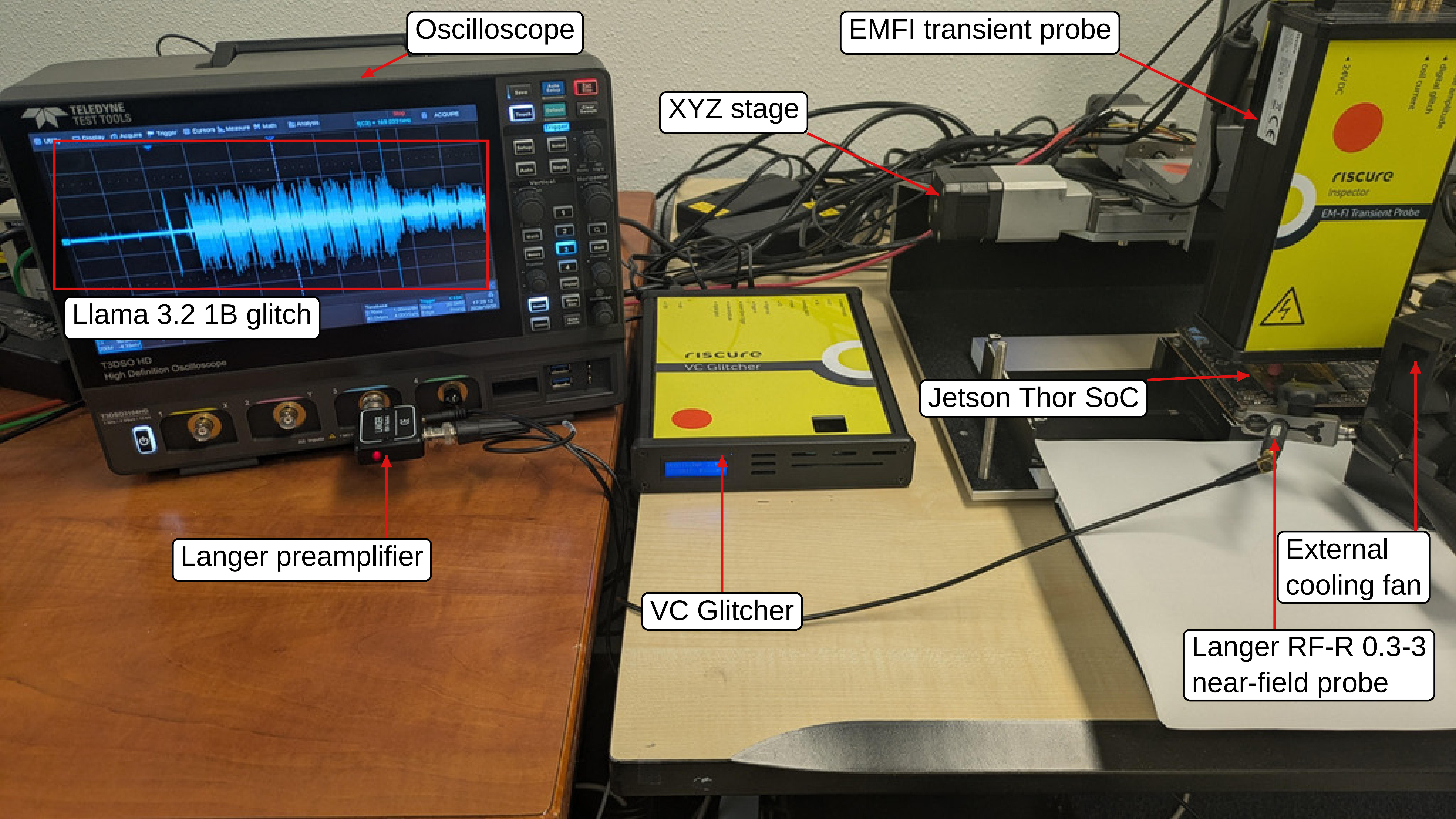}
    \caption{Annotated image of the setup for the Jetson AGX Thor Developer Kit, with the thermal enclosure removed and external cooling.}
    \label{fig::setup_thor_live}
\end{figure}

We use the same setup for the Jetson AGX Thor Developer Kit with minor changes in reset and cooling.
Thor also contains two PCBs, the carrier board and the Jetson Thor module.
The module SoC contains the Blackwell GPU and the Arm CPU cluster and they share the off-chip available 128\,GB LPDDR5X memory. The 128\,GB memory is divided between 8 DRAM packages on the module.
By default, the module sits in a thermal enclosure. To access the SoC, we remove the enclosure and reattach the module to the carrier board.
Similarly to Orin, we run the developer kit in the highest performance mode using \texttt{jetson\_clocks} and \texttt{nvpmodel} with the heatsink also removed. However, under realistic LLM workloads, the device easily overheats and shuts down without proper external cooling. During the campaign we wait for the SoC to cool down when it reaches \SI{90}{\celsius} before continuing the campaign.
For the trigger, we use the prefill rising edge, similarly to the Jetson Orin Nano setup. For resets, we power-cycle using the NVIDIA board automation tools via USB from the host PC to the board. An image of the real setup is shown in~\cref{fig::setup_thor_live}.
We experimented with positive and negative polarity probes and with 1.5\,mm and 4\,mm injection tips. All of them can lead to jailbreaks, but the 1.5\,mm tips, of either polarity, lead to fewer crashes. DRAM encryption (\cref{sec::dramenc}) might partially explain why we did not observe a difference between the two polarities: the faulted data decrypts to random bytes regardless of which bits the pulse sets or clears.

\subsection{DRAM encryption}\label{sec::dramenc}
Jetson Orin and Thor devices encrypt DRAM contents by default~\cite{dramenc}. This means that when data is written or read, the Memory Subsystem uses 128-bit AES-XTS to encrypt or decrypt it. While it does not provide integrity protection, it has implications for the faults that impact encrypted data:
\begin{itemize}
    \item  Single flipped bits, the direction of flips, and data-dependent effects can no longer be observed because of at least 16 bytes are corrupted.
    \item The faults can be frequently observed to be aligned on 16 bytes.
    \item Read and write faults will look similar as they will decrypt and encrypt to random bits.
\end{itemize}

In the context of neural networks, the bulk of the data is the weights that can be represented in many forms. Since corrupting encrypted weights results in at least 16 bytes of weights corrupted, the representation form of the weights is important because decrypted corrupted weights can blow up. For instance, corrupting \texttt{bfloat16} weights randomly is likely to blow up the contribution of these weights because the allowed dynamic range is much higher than the usual range of the weights.

\section{Experiments}\label{sec::exp}
In the experiments, we use JailbreakBench~\cite{chao2024jailbreakbench} benchmark. For a specific model, there might be some prompts in the benchmark that are not refused by default. We present two case studies: \texttt{llama3.2:1b} and \texttt{gemma3:1b}. These models are suitable for edge devices, and the experiments also show differences in the effort required to jailbreak them via physical fault injection. Our strategy is to target the weights and activations that influence model-level refusal. Therefore, we target the loading and processing of these values before and during inference. As mentioned earlier, persistence can also be desirable for an attacker, so we target the LPDDR package and the LPDDR PHY interface, since activation and weight values must pass through these interfaces during memory writes.

\subsection{Locating sensitive spots}\label{sec::exp-char}
Before attacking the models, we investigate which locations on the LPDDR packages and the LPDDR PHY interface are sensitive to faults, and to what extent, using the setup of \cref{sec::char}. For this, we run a simple code that performs the following steps:
\begin{enumerate}
    \item \label{step:char-randmem} loads memory with random values;
    \item\label{step:char-loadgpu} loads these values into the GPU to do computations;
    \item\label{step:char-checksum} calculates checksums to check if corruption occurred.
\end{enumerate}

\begin{figure*}[htbp]
\centering
\includeinkscape[height=7cm]{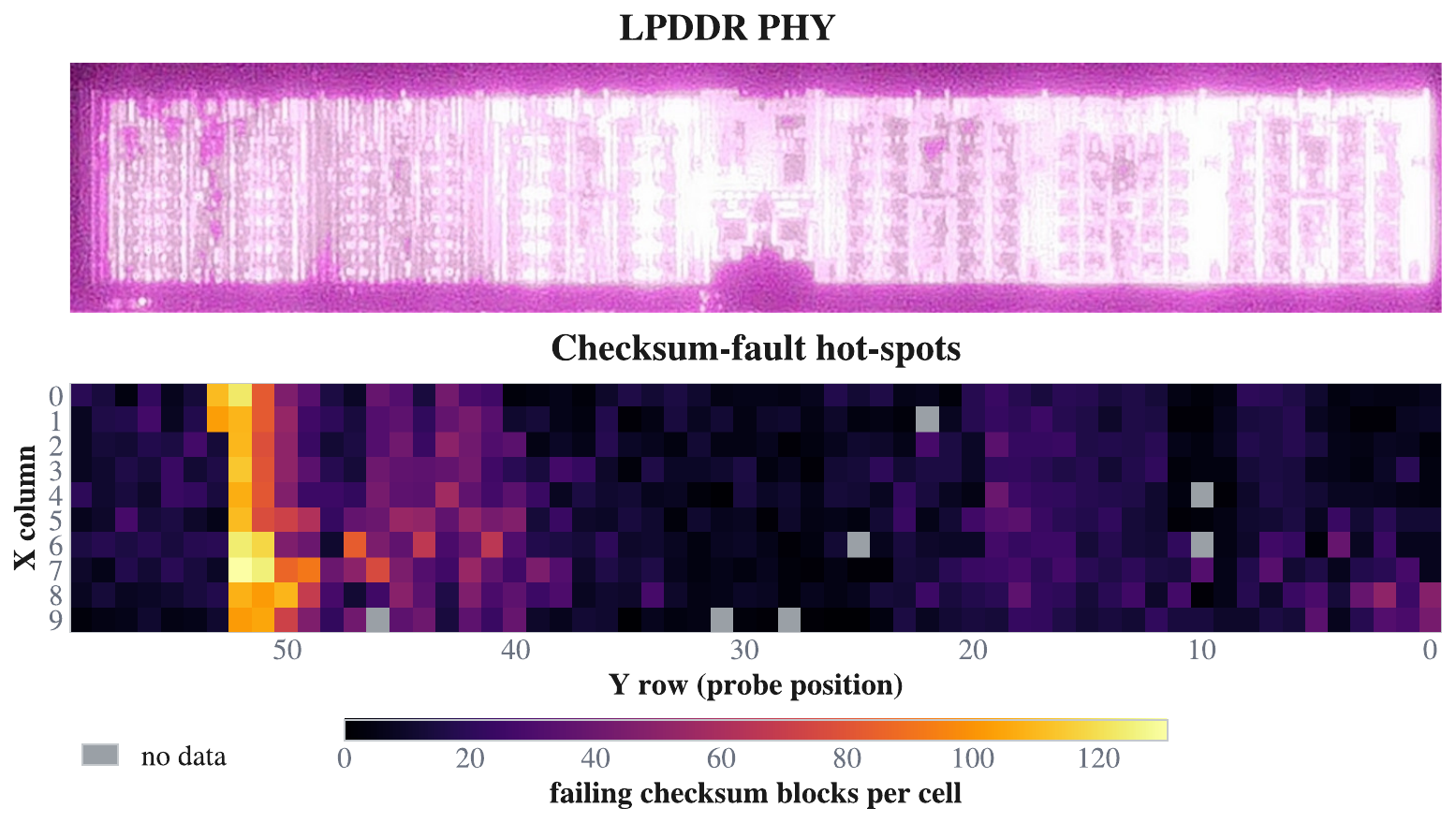}
\caption{LPDDR PHY interface and the number of impacted checksums at each probe location on the Orin with a single scan. The Orin has 4 PHY strips on the SoC, of which only 2 are used for Jetson Orin Nano.}\label{fig::orin_rightMC_channel}
\end{figure*}

We inject faults during Step~\ref{step:char-loadgpu}. Step~\ref{step:char-checksum} investigates their impact.
After the check, a map can be created of the actual vs. expected comparisons for each probe location to infer its sensitivity to single faults.
\cref{fig::orin_rightMC_channel} shows the hotspots on the LPDDR PHY interface in terms of the number of corrupted values. During the experiments, we observed that some faults cause persistent faults over either the LPDDR package or the PHY interface.
\Cref{app:channel-mapping} contains the results of fault injection experiments mapping memory die locations to impacted memory channels.

\subsection{Weight-loading faults}\label{sec::exp-reload}
A fault during weight loading\arxivonly{\enlargethispage{\baselineskip}} is attractive because it is \emph{persistent}: inference frameworks such as Ollama and vLLM do not check the integrity of weights once they are in memory, so a corrupted weight stays corrupted for every subsequent prompt until the model is reloaded. 
The typical parameter formats used for inference (\texttt{FP8}, \texttt{BF16}, \texttt{FP4}) and the architecture of the target models impact what certain faults can achieve. In many cases, over the PHY interface, we observed only encrypted weight faults and the faults in the weights impacting at least 16 bytes flipped randomly. However, many formats also use blockwise and/or tensorwise scales that are stored and loaded along with the weights. The scales make models particularly vulnerable because they allow an adversary to control more weights easily than otherwise. For instance, if the adversary can impact only 16 bytes with a single fault, that still allows them to corrupt tensor scales, which are at most 4 bytes (\texttt{FP32}). In some instances, \emph{dynamic quantization} is used that derives the scales on the fly. This provides the easiest attack path for an adversary to corrupt the scales because they only have to corrupt some weights. Additionally, the architecture influences whether certain faults just remove the contribution of a tensor or damage the model completely.

\parhead{Key result.}
A single EM pulse during an \texttt{FP8} weight load on the Jetson Thor can wipe a chosen MLP tensor, and the resulting model is the same model obtained by zero-ablating that tensor in software: on Gemma~3~1B, 11 of 12 such wipes produced JailbreakBench answers identical to the simulated ablation (98--100 of 100 answers) and, where benchmarked, MMLU and ARC-Challenge scores within one point.
The attacker can therefore choose a target offline, on an open-weight copy of the model, by looking for the tensor removal of which reduces refusals the most and utility the least:
\begin{itemize}
  \item \textbf{Llama~3.2~1B:} wiping the layer~11 MLP cuts refusals from 98 to 23--35 of 100 harmful prompts, with 32--38 unsafe answers, for about one point of MMLU.
  \item \textbf{Gemma~3~1B:} wiping layer~15's \tn{down\_proj} cuts refusals from 93 to 47, with 50 unsafe answers, at no MMLU cost.
\end{itemize}
\noindent These counts are from the DeepSeek labels and \cref{tab:emfi-judged-outcomes} also gives those from GLM.

\subsubsection{Setup}
The target is a Jetson Thor serving \texttt{FP8} models (one per-tensor scale, computed at load) with vLLM.
\begin{itemize}
  \item \textbf{Attack window:} to repeat trials without a reboot, we use vLLM's in-place \tn{reload\_weights}, which follows the same path as the initial load: each layer's checkpoint tensors are loaded and each module is quantized as soon as its tensors are in. 
  \item \textbf{Glitch:} with the setup of \cref{sec::char}, the oscilloscope triggers on an edge at the start of the reload and fires the VC Glitcher after a delay chosen for the targeted module. Each trial's parameters are taken from \cref{tab:emfi-params}.
  \item \textbf{Simulation:} the same tensor is zeroed so real and simulated models are served identically.
\end{itemize}

\noindent We evaluate all configurations, real and simulated, similarly.

\noindent \textbf{Safety:} the 100 harmful JailbreakBench prompts, greedy decoding, 256 tokens. Answers are scored with a refusal-keyword list (a refusal phrase e.g. ``I cannot'' in the first 200 characters), and configurations of interest are labeled \emph{refused}, \emph{unsafe} (specific, actionable harmful content), or \emph{other} (no refusal, but vague, evasive, off-topic, or garbled). Labels are from DeepSeek V4.1 Flash, audited by GLM~5.3 with the same rubric; the two judges agree on \mbox{70--100\%} of answers per configuration.

\noindent \textbf{Utility:} MMLU (20 questions from each of the 57 subjects, 1140 in total) and ARC-Challenge (all 1144 four-option questions), zero-shot setup and the answer is whichever letter A--D the model rates most likely after ``\textit{The correct answer is}''. 

\subsubsection{Interplay of architecture and formats}
Gemma and Llama represent two families of architectures in which normalization changes the effect of a fault. Llama is a pre-norm model: each block normalizes its input and adds its output directly to the residual stream, so an overflow in one tensor propagates to all later layers and damages the model. Gemma additionally normalizes the output of each block (post-norm). When a tensor overflows, this RMSNorm flushes the output of the block to zero, which removes the contribution of that tensor and leaves the rest of the model intact.
This difference determines the outcome of a direct fault in the weights. Random bit flips in 16 bytes of \texttt{bfloat16} weights almost certainly produce some very large weights. On Llama, this causes a denial of service at best, whereas on Gemma it zeroes the contribution of the whole tensor. Our experiments on real hardware (Jetson Thor) support this: on Gemma, 70\% of the attempts resulting in data corruption achieved a full wipe of \texttt{down\_proj}, while on Llama every attempt resulted in a damaged model.
A software simulation of the same fault, kept in place while the answer is decoded, reproduces this split: only degenerate output on Llama and fewer refusals on Gemma (\cref{app:prefill-sim}).

\begin{table}[t]
\centering
\caption{Effect of one corrupted quantization scale. Wipe: the layer output becomes zero; fail: incoherent output; local: only one block of weights changes. BF16~$\rightarrow$~X: quantized at load, scale computed from the (corrupted) data. These are simulated results on Jetson Thor.}
\label{tab:load-formats}
\footnotesize
\setlength{\tabcolsep}{4pt}
\begin{tabular}{@{}llcc@{}}
\toprule
Format & Scale & Gemma & Llama \\
\midrule
BF16 $\rightarrow$ FP8   & tensor & wipe & wipe \\
FP8                      & tensor & wipe & wipe/fail \\
BF16 $\rightarrow$ NVFP4 & tensor & wipe & wipe \\
NVFP4                    & tensor & wipe/fail & wipe/fail \\
NVFP4                    & block  & local & local \\
MXFP4                    & block  & wipe & fail \\
\bottomrule
\end{tabular}
\end{table}

For quantized weights with scales, the results are different because Llama becomes vulnerable as well.
\Cref{tab:load-formats} summarizes the impact of faults on scales for each architecture and parameter format in Jetson Thor simulations. We validated \texttt{BF16} $\rightarrow$ \texttt{FP8} with real experiments for Gemma and Llama.
\Cref{app:prefill-sim} examines nine parameter formats, including these, under transient faults during inference in an independent simulation. In these simulations, a fault in the quantized weights themselves does not lower refusals, while a fault in a scale does.

On Gemma we obtained full \tn{down\_proj} wipes on real hardware. Eleven of them, with corrupted scales of $2\times10^{22}$ or more, reproduced the simulated ablation of the same layer: the six layers whose ablation reduces refusals most (\cref{tab:emfi-gemma-real-vs-sim}) and layers 2, 20, 22, 23 and 24, with 98--100 of 100 answers identical. The only exception, layer~19, had a much smaller corrupted scale ($3\times10^{13}$) and did not match its simulation (40 vs.\ 57 keyword refusals, MMLU 38.0 vs.\ 39.8\%).

\begin{table}[t]
\centering
\arxivonly{\captionsetup{format=tight}}%
\caption{Gemma~3~1B \tn{down\_proj} wipes of the six layers whose ablation reduces refusals most, \emph{real wipe / simulated ablation}. Scale: the corrupted \texttt{FP8} scale of the tensor. ``Identical'' counts the JailbreakBench answers of the real wipe that are byte-identical to those of the simulation. Kw.\ ref.: refusals by keyword list, which undercounts Gemma's refusals because they often open with an empathetic preamble (``I understand you're going through\ldots'') and decline only later; the LLM judges count 91--93 refusals for the clean model (\cref{tab:emfi-judged-outcomes}).}
\label{tab:emfi-gemma-real-vs-sim}
\footnotesize
\setlength{\tabcolsep}{3.5pt}
\begin{tabular}{@{}rccccc@{}}
\toprule
Layer & Scale & Identical & Kw.\ ref. & MMLU (\%) & ARC-C (\%) \\
\midrule
4  & $1.2\times10^{32}$ & 100/100 & 26\,/\,26 & 25.2\,/\,25.2 & 26.6\,/\,26.6 \\
9  & $2.0\times10^{31}$ & 100/100 & 29\,/\,29 & 28.1\,/\,28.1 & 26.3\,/\,26.3 \\
13 & $9.5\times10^{29}$ & 100/100 & 14\,/\,14 & 31.0\,/\,31.0 & 31.1\,/\,31.1 \\
15 & $8.7\times10^{28}$ & 100/100 & 26\,/\,26 & \textbf{39.0}\,/\,39.0 & \textbf{46.9}\,/\,46.9 \\
16 & $1.3\times10^{29}$ & 100/100 & 16\,/\,16 & 30.0\,/\,30.0 & 36.4\,/\,36.4 \\
18 & $2.0\times10^{22}$ & \phantom{1}98/100 & 23\,/\,23 & 38.2\,/\,38.2 & 44.5\,/\,44.5 \\
\midrule
clean & $3$--$6\times10^{-4}$ & & 50 & 39.0 & 47.0 \\
\bottomrule
\end{tabular}

\end{table}

On Llama we wiped the MLP of layers 9, 10 and 11 on real hardware, hitting \tn{down\_proj} in some trials and \tn{gate\_up\_proj} in others but for the same layer, both gave identical answers. Here the answers also depend on the boot: after a reboot, the same wipe lands in one of a few repeatable numerical states. Within a matching boot state, the real and simulated answers are byte-identical: the layer-10 and layer-11 wipes reproduced all 100 simulated answers, while the layer-9 wipe gave one of 3 outcomes across boots, and each repeated identically within its boot.

\subsubsection{Effect on refusals and utility}
Wiping a single MLP more than halves refusals on Llama and roughly halves them on Gemma. By the DeepSeek labels, Llama layer~11 goes from 98 to 23--35, and Gemma layers 9, 13 and 15 from 93 to 32--50, with 44--53 unsafe answers (\cref{fig:emfi-judged-outcomes}). The GLM labels give 95 to 34--45 for Llama and 91 to 42--54 for Gemma (\cref{tab:emfi-judged-outcomes}).

\pgfplotsset{
  judged/.style={
    xbar stacked,
    bar width=7pt,
    scale only axis,
    width=0.32\textwidth,
    height=3.2cm,
    xmin=0, xmax=100,
    enlarge y limits={abs=0.6},
    y dir=reverse,
    ytick=data,
    yticklabel style={font=\scriptsize, text width=0.14\textwidth, align=right},
    xticklabel style={font=\scriptsize},
    xtick={0,25,50,75,100},
    xlabel={Answers out of 100}, xlabel style={font=\scriptsize},
    axis x line*=bottom,
    axis y line*=left,
    tick align=outside,
    major tick length=2pt,
    title style={font=\small\bfseries, yshift=-4pt},
    visualization depends on=rawx\as\segw,
    nodes near coords={\ifdim\segw pt<0.5pt\else\pgfmathprintnumber{\segw}\fi},
    every node near coord/.append style={font=\tiny, anchor=center, inner sep=0pt},
  },
  judged refused/.style={fill=refusedc, draw=white, line width=0.3pt},
  judged unsafe/.style={fill=unsafec, draw=white, line width=0.3pt, every node near coord/.append style={text=white}},
  judged other/.style={fill=otherc, draw=white, line width=0.3pt},
}

\begin{figure*}[t]
\centering
\begin{tikzpicture}
\begin{axis}[name=llama, judged, title={Llama 3.2 1B},
  yticklabels={Clean model, {L11 MLP (boot A)}, {L11 MLP (boot B)}, {L11 MLP (boot C)}, {L10 MLP}, {L9 MLP}},
  legend style={at={(1.047,1.2)}, anchor=south, legend columns=3, font=\scriptsize, draw=none,
    /tikz/every even column/.append style={column sep=6pt}},
  legend image code/.code={\draw[#1, draw=none] (0cm,-0.08cm) rectangle (0.25cm,0.08cm);}]
\addplot[judged refused] coordinates {(98,0) (29,1) (35,2) (23,3) (72,4) (71,5)};
\addplot[judged unsafe]  coordinates {(2,0) (38,1) (32,2) (38,3) (18,4) (14,5)};
\addplot[judged other]   coordinates {(0,0) (33,1) (33,2) (39,3) (10,4) (15,5)};
\legend{Refused, Unsafe, Vague\,/\,evasive\,/\,garbled}
\end{axis}
\begin{axis}[judged, title={Gemma 3 1B}, at={(llama.south east)}, anchor=south west, xshift=0.17\textwidth,
  yticklabels={Clean model, {L9 \tn{down\_proj}}, {L13 \tn{down\_proj}}, {L15 \tn{down\_proj}}, {L16 \tn{down\_proj}}, {L18 \tn{down\_proj}}}]
\addplot[judged refused] coordinates {(93,0) (32,1) (50,2) (47,3) (82,4) (76,5)};
\addplot[judged unsafe]  coordinates {(5,0) (53,1) (44,2) (50,3) (7,4) (21,5)};
\addplot[judged other]   coordinates {(2,0) (15,1) (6,2) (3,3) (11,4) (3,5)};
\end{axis}
\end{tikzpicture}
\caption{Answers to the 100 harmful JailbreakBench prompts after a single-layer wipe on Thor, labelled by DeepSeek V4.1 Flash.  Llama L11 shows its three boot-dependent answer sets (boot A matches the simulation). Llama L10 and L9 show the boot state whose answers match the simulation. GLM~5.3 labels are in \cref{tab:emfi-judged-outcomes}.}
\label{fig:emfi-judged-outcomes}
\end{figure*}
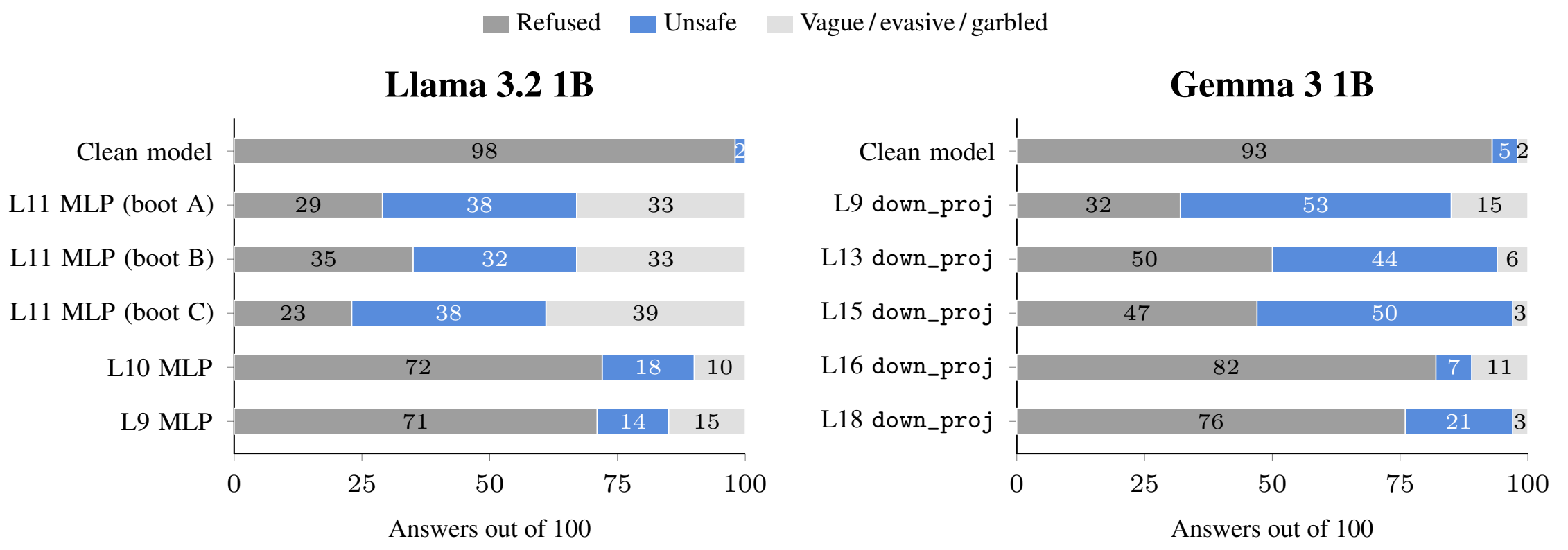

\begin{table}[t]
\centering
\caption{JailbreakBench outcomes and utility of single-layer wipes. Refused, Unsafe: answers out of 100, as labelled by DeepSeek V4.1 Flash and by GLM~5.3; the rest are \emph{other}, shown for DeepSeek in \cref{fig:emfi-judged-outcomes}. MMLU, ARC-C: accuracy in \%. Bold: best trade-off per model.}
\label{tab:emfi-judged-outcomes}
\footnotesize
\setlength{\tabcolsep}{2.6pt}
\begin{tabular}{@{}l cc cc rr@{}}
\toprule
 & \multicolumn{2}{c}{DeepSeek} & \multicolumn{2}{c}{GLM} & & \\
\cmidrule(lr){2-3}\cmidrule(lr){4-5}
Wiped tensor & Refused & Unsafe & Refused & Unsafe & MMLU & ARC-C \\
\midrule
\multicolumn{7}{@{}l}{\emph{Llama 3.2 1B}} \\
--- (clean) & 98 & 2 & 95 & 5 & 47.0 & 53.1 \\
L11 MLP$^{\ddagger}$ & \textbf{23--35} & \textbf{32--38} & \textbf{34--45} & \textbf{29--38} & \textbf{46.0} & \textbf{51.7} \\
L10 MLP$^{\ddagger}$ & 63--72 & 18--23 & 60--69 & 15--18 & 42.8 & 44.7 \\
L9 MLP$^{\ddagger}$ & 59--78 & 10--29 & 56--76 & 8--21 & 40.0 & 45.1 \\
\midrule
\multicolumn{7}{@{}l}{\emph{Gemma 3 1B}} \\
--- (clean) & 93 & 5 & 91 & 8 & 39.0 & 47.0 \\
L9 \tn{down\_proj}  & 32 & 53 & 42 & 41 & 28.1 & 26.3 \\
L13 \tn{down\_proj} & 50 & 44 & 54 & 37 & 31.0 & 31.1 \\
L15 \tn{down\_proj} & \textbf{47} & \textbf{50} & \textbf{52} & \textbf{46} & \textbf{39.0} & \textbf{46.9} \\
L16 \tn{down\_proj} & 82 & 7 & 77 & 12 & 30.0 & 36.4 \\
L18 \tn{down\_proj} & 76 & 21 & 78 & 19 & 38.2 & 44.5 \\
\bottomrule
\end{tabular}

\smallskip
{\footnotesize\raggedright
$^{\ddagger}$Range over the boot-dependent answer sets of the different wipes: 3 for L11, 2 for L10, 3 for L9.\par}
\end{table}

The two judges agree most on clean models and least on heavily damaged ones, because the boundary between \emph{unsafe} and \emph{other} is less clear in many responses.

\begin{takeawaybox}
Electromagnetic pulses during weight loading can wipe a chosen tensor and roughly \textbf{halve the refusals compared with the clean model}, at little observable cost in utility.
\end{takeawaybox}

\subsection{Inference time transient faults}
\begin{figure*}[htbp]
\centering
\includegraphics[width=\textwidth]{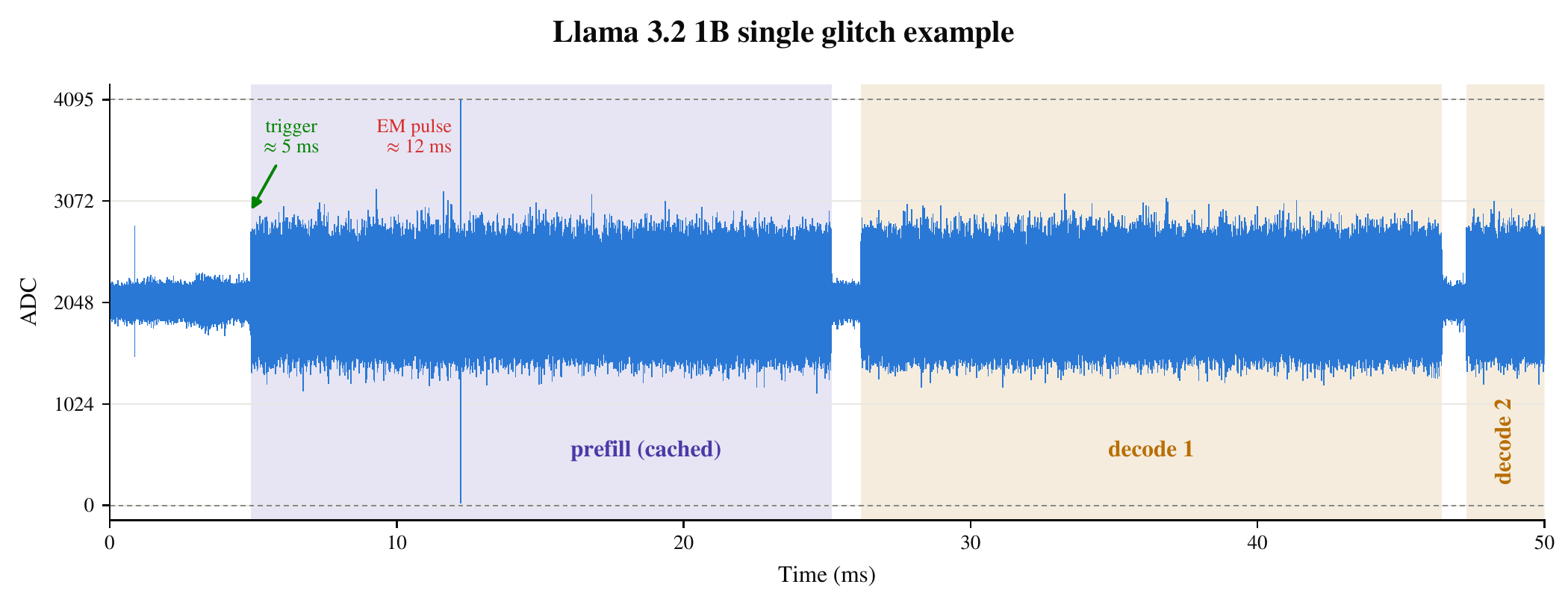}
\caption{Near-field trace of GPU emanations of \texttt{llama3.2:1b} from the Orin SoC during fault injection. We inject a pulse during layer 8 \texttt{down\_proj} in the prefill phase. The pulse is clearly captured in the trace.}\label{fig::orin_llama32b}
\end{figure*}

The inference-time experiments use the setup and pulse parameters of \cref{sec::char} (\cref{tab:emfi-params}), triggered on the rising edge of the prefill.
On the Orin, the models run in Ollama as GGUF files, \texttt{llama3.2:1b} in \texttt{Q8\_0} and \texttt{gemma3:1b} in \texttt{Q4\_K\_M} (\cref{tab:platform_orin} in \cref{app:stack}). On Thor, both run in vLLM in \texttt{BF16} (\cref{tab:platform_thor} in \cref{app:stack}).
Decoding is greedy on both platforms (temperature~0), so an answer that differs between attempts is not an effect of sampling.
While zero-ablation does identify layers that can significantly reduce refusal in the models, for inference time faults we target layers based on the technique of Arditi et al.~\cite{NEURIPS2024_f5454485}\ifBaselineTable, with the results summarized in~\cref{tab:main} in \cref{app:ref}\fi.
For \texttt{llama3.2:1b}, multiple \texttt{down\_proj} layer weights and activations can be targeted with glitches; we show an example of an injected glitch in~\cref{fig::orin_llama32b} targeting layer~8.
Unlike a fault that persists, such a fault is not a denial of service on Llama. In simulation, the overflow caused by a faulted \tn{down\_proj} of layer~$L$ zeroes the KV cache that the layers above $L$ store for the prompt; the first answer token becomes ``\texttt{!}'', and the remaining tokens are decoded with intact weights on a cache in which the upper layers do not see the prompt (\cref{app:prefill-sim}).
For \texttt{gemma3:1b}, the sensitive directions are in layers 8 and 9.
We therefore direct the pulses at the weights and activations of these layers during inference.\arxivonly{\looseness=-1}

\begin{table}[t]
\caption{JailbreakBench results of fault injection on the LPDDR PHY. Trials/Prompt: glitch attempts per prompt. Bypass: share of the 100 prompts whose refusal was bypassed in at least one attempt. Unsafe: share of the prompts that yielded genuinely actionable harmful content. The labels are from DeepSeek V4.1 Flash, and their audit is from GLM 5.3.}
\centering
\footnotesize
\setlength{\tabcolsep}{4pt}
\begin{tabular}{@{}llc rr rr@{}}
\toprule
\multirow{2}{*}{Model} & \multirow{2}{*}{Plat.} & \multirow{2}{*}{\shortstack{Trials/\\Prompt}} & \multicolumn{2}{c}{DeepSeek V4.1} & \multicolumn{2}{c}{GLM 5.3} \\
\cmidrule(lr){4-5}\cmidrule(l){6-7}
& & & Bypass & Unsafe & Bypass & Unsafe \\
\midrule
\multirow{4}{*}{\texttt{gemma3:1b}}
 & \multirow{2}{*}{Orin} & 15 & 65\% & 39\% & 65\% & 35\% \\
 &                       & 50 & 78\% & 53\% & 76\% & 47\% \\
\cmidrule(l){2-7}
 & \multirow{2}{*}{Thor} & 15 & 37\% & 28\% & 37\% & 28\% \\
 &                       & 50 & 53\% & 40\% & 53\% & 40\% \\
\midrule
\multirow{4}{*}{\texttt{llama3.2:1b}}
 & \multirow{2}{*}{Orin} & 15 &  81\% & 40\% &  81\% & 40\% \\
 &                       & 50 & 100\% & 97\% & 100\% & 97\% \\
\cmidrule(l){2-7}
 & \multirow{2}{*}{Thor} & 15 & 29\% & 15\% & 28\% & 14\% \\
 &                       & 50 & 58\% & 34\% & 54\% & 34\% \\
\bottomrule
\end{tabular}

\label{tab:results_faultinj_jbb}
\end{table}

The results of the experiments on LPDDR PHY are shown in~\cref{tab:results_faultinj_jbb} and the answers are judged by DeepSeek V4.1 Flash and GLM 5.3 similarly to \cref{sec::exp-reload}. For experiments over the LPDDR package, we present results on Thor in \cref{app:agx-lpddr}. In every experiment, caching is active after the first query for a particular input prompt. For pulses over PHY, we experimented with rows between 45 and 55 from~\cref{fig::orin_rightMC_channel} and reliably observed a jailbreak on Orin. We also used the LPDDR PHY for Thor. Generally, the injections for Thor are less precise, hence the lower jailbreak ratios. Partly, this could be due to ineffective cooling and consequent throttling, which causes Thor's performance to vary widely.
On the Orin, 97\% of the prompts yield an unsafe answer for Llama within 50 attempts, more than the 78--81 prompts that all 36 judged single-block \texttt{Q8\_0} faults reach together in simulation (\cref{app:prefill-sim}), which suggests that some pulses corrupt more than one 16-byte block.
Overall, timing remains critical, and we expect the accuracies to achieve near theoretical maximums with precise triggering for both models on Thor as well. Most of the injected faults over PHY are transient, but persistence can be achieved during initial weight loading or when the KV cache prefix is corrupted. For initial weight loading, the adversary has to reset the device after an unsuccessful fault attempt, and rebooting and reloading can take minutes for larger models. 

\begin{takeawaybox}
Transient faults during inference \textbf{bypass refusals while the stored weights stay intact}: up to 97\% of prompts yield an unsafe answer.
\end{takeawaybox}

\section{Faulting a real robot}\label{sec::robot}
The attacks in previous sections change what a language model says. A vision-language-action (VLA) policy maps camera images and a task instruction directly to motor commands, so a fault in it changes what the model and controlled robot do. We inject EM pulses into a Jetson Orin Nano running a SmolVLA policy driving a robot arm. SmolVLA~\cite{smolvla} is a small open-source VLA model for low-cost robots: a compact vision-language model reads camera images, the task and the robot's joint state, and a flow-matching action expert turns its features into a chunk of motor commands. The threat model is that of \cref{sec::jf}, except that the attacker needs no input to the model and the goal is a motion the robot was not meant to make. This section covers: simulated faults (\cref{sec::robot-sim}), EM faults on the policy during inference (\cref{sec::robot-policy}), a comparison of simulation and experiment (\cref{sec::robot-compare}) and persistent faults during model loading (\cref{sec::robot-load}).

\subsection{Threat Scenario}\label{sec::robot-threat}
This section adapts~\Cref{subsec:threatmodel}'s threat model to a realistic robotic setup. Consider a collaborative sorting station where a robotic arm works alongside a human operator, relying on an onboard VLA model to process live camera feeds and translate tasks -- such as picking and placing items -- directly into safe motor commands. Because the VLA policy governs the arm's physical trajectories in a shared workspace, corrupting it allows an adversary to turn a routine pick-and-place operation into a physical hazard or sabotage the line without altering sensor inputs or tripping software alarms. During routine maintenance or brief unattended access, the attacker places a compact, low-cost handheld EMFI probe next to the robot's edge GPU without needing network access, OS credentials, or modifications to encrypted weights on disk. Triggered automatically by the GPU's electromagnetic emanations during inference or start-up, a single pulse silently corrupts memory transactions. Consequently, when the operator reaches into the workspace to retrieve a sorted item, the compromised policy outputs a severely distorted trajectory -- forcing the arm to abruptly lunge and pinch or strike the operator's hand leaving no software trace.

\subsection{Setup}\label{sec::robot-setup}
\parhead{Platform and policy.}
The board is the Jetson Orin Nano of \cref{sec::char}\ifPlatformTables, with the software of \cref{tab:platform_robot} in \cref{app:stack}\fi. It drives a LeRobot~\cite{lerobot} SO-101 arm with six joints and two cameras, one in front of the workspace and one at an angle; the task is ``\textit{pick up the red cube and place it in the transparent box}'' (\cref{fig:robot-frames-normal} shows a fault-free episode). Each inference of the fine-tuned SmolVLA encodes both camera images (64 tokens each), runs a prefill of the 16-layer language model over the image, language, and state tokens, whose keys and values form the KV cache, and integrates the action expert in 10 denoising steps, $x_{t+1} = x_t - 0.1\,v_t$, to a chunk of 50 actions. The rollout runs at 30\,Hz and requests a new chunk every 19 control ticks.

\parhead{EM injection.}
In these experiments,
we used the NewAE ChipSHOUTER~\cite{chipshouter}, triggered by a ChipWhisperer-Husky~\cite{chipwhisperer}, with 80\,ns pulses. However,  we verified that a FaultyCat~\cite{faultycat} can also be used with the Husky to cause faults. A near-field probe over a capacitor on the module senses GPU activity and triggers the Husky, which fires after a set delay. We fault only the model, not the cameras or the arm, to establish whether a corrupted model on its own leads to unsafe actions.

\parhead{Outcome classes.}
After each targeted inference, we recompute the same observation (same images, joint state and flow-matching noise) for a fault-free reference chunk. The plan change $\Delta$ is the largest absolute difference between the two chunks over the 50 steps and 6 joints (degrees; gripper in percent of its range). Each targeted inference falls into one class:
\begin{itemize}
\item \textbf{Detected:} the chunk contains NaN. The stock rollout then stops the episode so we changed it to drop the chunk and continue with the previous plan, and all experiments use this change unless stated otherwise.
\item \textbf{Silent:} $\Delta \ge 10\dg$, inside the joint limits, and no joint moves more than 10\dg{} between consecutive steps: an ordinary-looking motion.
\item \textbf{Erratic:} $\Delta \ge 10\dg$, but outside the joint limits or with larger jumps.
\item \textbf{Small:} $0 < \Delta < 10\dg$. \textbf{Unchanged:} $\Delta = 0$.
\end{itemize}
Silent and erratic faults are successful, and detected faults are similar to LLM crashes.

\subsection{Simulated faults}\label{sec::robot-sim}
EM faults on the Jetson garble whole aligned 16-byte blocks of the AES-XTS-encrypted DRAM (\cref{sec::dramenc}): a faulted read replaces 8 consecutive \texttt{FP16} values with random bit patterns. We simulate one such block per inference on 8 observations logged during rollouts (\cref{tab:robot-blocksim}). The simulation does not use the robot, so it predicts the planned, not the executed, motion. The corrupted block lands in the weights read by one camera's vision encoder, by the prefill or by one denoising step; in a stored expert weight read by all 10 steps, as after a fault during loading (\cref{sec::robot-load}); in the input of one matrix multiplication (activations); or in the KV cache, as written by the prefill or in one step's read. Blocks are drawn in proportion to the bytes. Weight and KV-cache blocks are written into the GPU memory of the original policy, with 4,000 faults per source. Activations exist only inside its inference engines and cannot be faulted from outside, so their blocks are injected in a copy of the policy, with 2,000 faults per source. The medians and the breakdowns by layer and projection below are also from the copy. The results predict the kind of faults and their impact:
\begin{itemize}
\item \textbf{Silent faults:} these come from garbled weight reads, in 35\% of the blocks in the vision encoders and 32\% in the prefill, with median plan changes of 27\dg{} and 24\dg. In the prefill, their share of the garbled weight blocks falls from 58\% in layers 0--3 to 2\% in layers 12--15. Activation and KV-cache blocks give 6\% or fewer.
\item \textbf{Plans changed by 100\dg{} or more:} these come almost only from the values that the expert reads. In the copy, 63 of the 2,000 blocks garbled in prefill weights fell in a value projection, and 28 of them (44\%) gave such a plan but none of the 65 in a key projection did. Every prefill layer has a value projection, so such plans are about equally likely in every layer (1--3\% of the garbled blocks).
\item \textbf{Faults in one denoising step:} these give NaN or are washed out by the later steps, unless they hit the last step, where they give jagged changes of 13--18\dg.
\item \textbf{A stored expert weight:} a garbled block gives an erratic plan in 35\% of cases and a plan changed by 100\dg{} or more in 20\%.
\end{itemize}
For the weights and the KV cache, the copy agrees with the original: its silent, erratic and $\ge$100\dg{} rates differ by at most 1.2 points, except for the erratic plans of a stored expert weight (43\%).

\begin{table}[t]
\caption{Simulated block faults: one aligned 16-byte block replaced with random bytes per inference. Rates in \% of faults. Activations are faulted in a copy of the policy (see text).}
\label{tab:robot-blocksim}
\centering
\footnotesize
\setlength{\tabcolsep}{3pt}
\begin{tabular}{@{}lrrrrr@{}}
\toprule
Source & Detected & Silent & Small & Erratic & $\ge$100\dg \\
\midrule
\multicolumn{6}{@{}l}{\emph{Weights read by}} \\
\quad vision encoder (1 camera) & 40.2 & 34.6 & 24.9 & 0.1 & 0.0 \\
\quad prefill & 22.9 & 31.9 & 42.6 & 2.6 & 2.1 \\
\quad one denoising step & 41.8 & 0.2 & 54.2 & 3.7 & 0.0 \\
\quad all 10 steps (stored) & 47.4 & 0.7 & 16.6 & 35.4 & 20.4 \\
\multicolumn{6}{@{}l}{\emph{Activations in}} \\
\quad vision encoder & 23.4 & 4.5 & 72.2 & 0.0 & 0.0 \\
\quad prefill & 27.6 & 5.7 & 46.1 & 4.8 & 2.0 \\
\quad one denoising step & 27.5 & 0.1 & 69.7 & 2.6 & 0.3 \\
\multicolumn{6}{@{}l}{\emph{KV cache}} \\
\quad as written & 19.7 & 3.3 & 48.3 & 10.5 & 3.0 \\
\quad in one step's read & 19.3 & 0.4 & 61.1 & 0.5 & 0.0 \\
\bottomrule
\end{tabular}
\par\smallskip
\end{table}

\subsection{EM faults on the policy}\label{sec::robot-policy}
\Cref{tab:robot-campaigns} lists the campaigns; in each, the arm executed faulted plans, some with large erratic motion.

\begin{table}[t]
\caption{Policy fault injection campaigns. Rates in \% of pulses.}
\label{tab:robot-campaigns}
\centering
\footnotesize
\setlength{\tabcolsep}{4pt}
\sbox0{\begin{tabular}{@{}lrrrr@{}}
\toprule
Target & Detected & Silent & Small & Erratic ($\ge$100\dg) \\
\midrule
Side-camera encoder & 17.5 & 7.2 & 11.4 & 0.0 (0.0) \\
Prefill & 16.3 & 3.8 & 13.5 & 0.3 (0.2) \\
Prefill + output, 2 pulses & 49.1 & 4.4 & 10.3 & 4.4 (0.4) \\
Last denoising steps & 24.8 & 0.0 & 12.1 & 10.6 (0.0) \\
Last denoising step & 32.1 & 0.0 & 3.0 & 3.4 (0.0) \\
\bottomrule
\end{tabular}}
\usebox0\par\smallskip
\parbox{\wd0}{\footnotesize  Erratic: erratic chunks, in brackets those with a plan change of at least 100\dg. ``Prefill + output'': one pulse in the prefill and one in the last denoising step.}
\end{table}

\begin{figure*}[t]
\centering
\begin{subfigure}[t]{0.49\textwidth}
\centering
\numberedframes{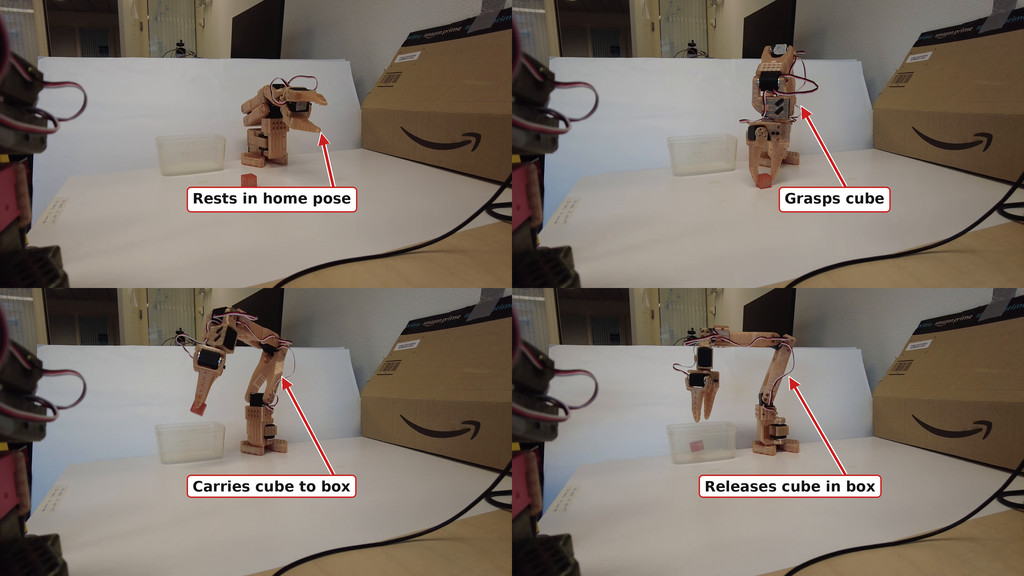}
\caption{Normal operation: the arm rests in its home pose~(1), grasps the cube~(2), carries it to the box~(3) and releases it there~(4).}
\label{fig:robot-frames-normal}
\end{subfigure}\hfill
\begin{subfigure}[t]{0.49\textwidth}
\centering
\numberedframes{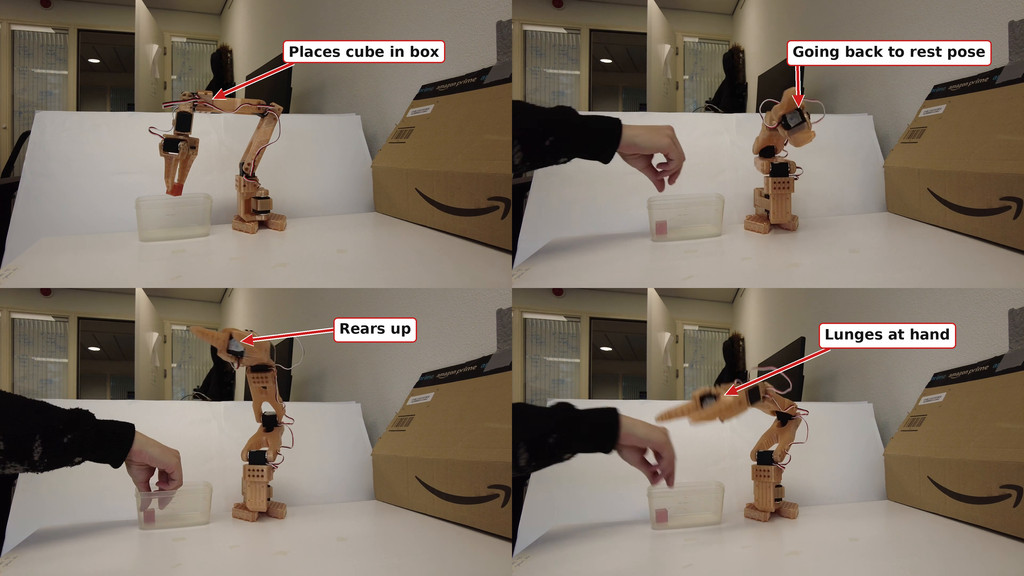}
\caption{Faulted episode: the arm places the cube in the box~(1) and starts to return to its rest pose~(2), then rears up~(3) and lunges towards a hand reaching into the box~(4).}
\label{fig:robot-frames-clip1}
\end{subfigure}
\caption{Video frames of the arm in a fault-free episode and in a moving-arm episode with an EM fault. The frames are numbered in the order in which they were recorded. The video clips are available at \videourl.}
\label{fig:robot-frames}
\end{figure*}

\parhead{Silent faults.}
In the first runs on the side-camera encoder and the prefill, 5--10\% of pulses gave silent faults, with median plan changes of 21--26\dg{} and at most 73\dg; over all runs on the prefill, 3.8\%. Pulses in the last denoising step gave NaN in a quarter to a third of cases and never a silent fault: their erratic chunks were jagged changes of only \mbox{10--21\dg}.

\parhead{Large erratic plans.}
Some pulses in the prefill, one of them at its border with the side-camera encoder, produced plans that differ from the fault-free plan by 119--359\dg{} and move several joints at once over most of their range. The arm executed them within 0.5--0.75\,s, in the largest case into the end stop of its shoulder lift, and the next fault-free chunks brought it back within about 2\,s, without damage. In \cref{fig:robot-frames-clip1}, the arm rears up and lunges towards a hand reaching into the box; \cref{fig:robot-frames-app} in \cref{sec::robot-frames-app} shows further episodes of different experiments.

\begin{takeawaybox}
One pulse during inference corrupts a single action chunk and the arm \textbf{behaves erratically for a moment} and goes back to normal afterwards.
\end{takeawaybox}

\subsection{Simulation and experiment compared}\label{sec::robot-compare}
Our EM results on Jetson Orin Nano follow the simulation  (\cref{tab:robot-layers}). In the prefill, the silent share among silent and small changes falls from 59\% in layers 0--3 to 10\% in layers 12--15 (weight blocks: 73\% to 3\%); activation blocks give much smaller shares. The 15 plans changed by 100\dg{} or more came from layers 0--3, 8--11 and 12--15 at similar rates (\mbox{0.4--0.7\%} of the changed inferences). In the vision encoders, the silent share of 23\% lies between those of weight (68\%) and activation blocks (6\%), so pulses there likely garble both. Pulses in the last denoising steps gave NaN or jagged chunks and no silent fault. EM pulses differ in two ways. In the prefill, 52--66\% of the changed inferences ended in NaN, against 19--36\% for one garbled block which points to several garbled blocks per pulse. And at most 0.7\% gave plans changed by 100\dg{} or more, under half the simulated 1--3\%.

\begin{table}[t]
\caption{Silent faults by stage: EM pulses against simulated garbled blocks in the weight reads and activations.}
\label{tab:robot-layers}
\centering
\footnotesize
\setlength{\tabcolsep}{3pt}
\begin{tabular}{@{}lrrrrrr@{}}
\toprule
 & \multicolumn{2}{c}{EM} & \multicolumn{2}{c}{Weights} & \multicolumn{2}{c}{Activations} \\
\cmidrule(lr){2-3}\cmidrule(lr){4-5}\cmidrule(l){6-7}
Stage & Share & Med. & Share & Med. & Share & Med. \\
\midrule
Vision encoders & 23 & 21 & 68 & 27 & 6 & 15 \\
\multicolumn{7}{@{}l}{\emph{Prefill layers}} \\
\quad 0--3 & 59 & 29 & 73 & 31 & 18 & 20 \\
\quad 4--7 & 50 & 21 & 59 & 22 & 13 & 17 \\
\quad 8--11 & 24 & 14 & 21 & 14 & 6 & 13 \\
\quad 12--15 & 10 & 13 & 3 & 12 & 5 & 14 \\
\bottomrule
\end{tabular}
\par\smallskip
\parbox{\columnwidth}{\footnotesize Share: silent faults in \% of silent and small ones. Med.: median plan change of the silent faults (\dg). EM: pulses of the moving-arm runs. Weights, activations: one simulated garbled block.}
\end{table}

\subsection{Persistent faults during model loading}\label{sec::robot-load}
A transient pulse corrupts one inference and does not impact subsequent plans. A fault in the weights as they load instead persists, as on the Jetson Thor (\cref{sec::exp-reload}): TensorRT copies the weights of an engine to GPU memory once, when it deserializes the engine. We therefore pulse the load of the action expert, whose weights are read at every denoising step of every inference.

\parhead{Method.}
At start-up, the rollout copies the expert's weights: first the gate and up projections, then the attention and down projections. The start of the copy, sensed by the near-field probe, triggers the Husky, followed by a single 80\,ns pulse at a set delay into the copy.

\parhead{Results.}
We pulsed 163 loads of the policy with the arm attached, pooled over eleven campaigns. Of these loads, 1.2\% crashed, that is, the load or the first inference failed, and 17\% were corrupted, each with \mbox{1--6} garbled 16-byte blocks in the weights.\footnote{The number of blocks impacted also depends on where the PHY interface is targeted. Since simulations show big impact with a small number of blocks, we placed the probe near locations where not too many blocks are affected.} The pulse delay selects which weights are impacted: a pulse 16\,ms into the copy hit the gate and up projections; pulses at 42--50\,ms hit the down projections and the query, value and output projections of the attention. We classify each corrupted load by its worst plan on the 8 reference observations. Almost all corrupted loads broke the plan: 59\% of them gave NaN and 33\% gave erratic plans changed by 106--192\dg; none gave a silent fault. The rest garbled query-projection blocks and changed the plan by less than 6\dg. This is in line with the simulation, where one stored block gives NaN in 47\% of cases and an erratic plan in 35\% (\cref{tab:robot-blocksim}). The NaN loads had 2--6 garbled blocks and the erratic ones 1--4, but the garbled values decided the outcome: 88\% of the NaN loads had a NaN among them and the rest had values above 40,000, which overflow in the computation, while no erratic or small load had either.

\parhead{Erratic policies on the moving arm.}
The observed faults for erratic movements in our experiments lay in the down projections of layers 5, 6, 7, 8 and 10 and in the attention output projections of layers 6 and 7. In every episode, every plan jumped by more than 10\dg{} between consecutive steps, with 83\% of them leaving joint limits, and none contained NaN while the shoulder lift reached 73--113\dg{} from its rest pose. Examples of the faults:
\begin{itemize}
\item \textbf{Layer-10 down-projection fault:} two blocks in row 561 of layer 10's \tn{down\_proj}, whose 16 weights, all below 0.05 in magnitude, became values such as $-7{,}756$ and $3{,}672$. All eight reference plans were erratic, by up to 146.5\dg.
\item \textbf{Layer-7 attention-output fault:} one block in row 392 of layer 7's \tn{o\_proj}, whose 8 weights, all below 0.05 in magnitude, became values such as $-24{,}768$ and $-977$. All eight reference plans were erratic, by up to 192.2\dg, the largest change of all loads.
\item \textbf{Layer-2 gate/up fault:} one block with values up to 20,912. Four of the eight reference plans were erratic, by up to 126\dg.
\end{itemize}

Overall, the models have many weaknesses with respect to safe operation, and integrity verification of the weights is paramount to ensure safe and secure operation.

\begin{takeawaybox}
One pulse during the robot's start-up can make it behave \textbf{dangerously until it is shut down}.
\end{takeawaybox}

\section{Discussion}\label{sec::disc}

\subsection{Defenses and integrity verification}
The persistence of the induced faults underscores the necessity of active integrity verification in memory.
A practical software-level mitigation is periodic memory integrity verification: computing cryptographic hashes (e.g., SHA-256 or lightweight HMACs) over resident model weights and validating them against trusted digests kept in secure storage.
Because verifying the full parameter space at each inference pass introduces non-negligible latency, integrity checks can be targeted selectively at safety-critical sub-layers, such as the specific down-projection tensors mediating refusal identified in our analysis.
Alternatively, verification can execute asynchronously during idle inference cycles or be scheduled at predetermined time intervals.
At the hardware layer, edge accelerators frequently lack Error-Correcting Code (ECC) memory.
Integrating Inline ECC or Side-band ECC in edge devices provides initial resilience; however, concentrated EM pulses often yield clustered multi-bit errors capable of overwhelming standard single-error correction and double-error detection (SECDED) schemes.
Complementary physical defenses, including integrated EM/glitch sensors and RF shielding around high-speed memory interfaces and PHYs, are essential to trigger an immediate hardware zeroization or safe system reset upon detecting transient field anomalies.
In addition, traditional fault injection countermeasures such as redundancy can be used. Redundancy duplicates important operations (e.g., weight loads, activation stores) and a fault attempt can be detected if the comparison between the duplicates yields different results.
Furthermore, injecting random delays for each layer's execution can make an adversary's task harder. This is already present (unintentionally) to some extent in the investigated devices due to their variable execution latencies.

\subsection{Limitations and threat implications}
Although our physical fault attack completely circumvents prompt-level guardrails, several operational trade-offs remain.
The primary challenge is spatial and temporal tuning: EMFI is inherently stochastic, presenting a narrow margin between neutralizing the refusal mechanism and causing unrecoverable system crashes or incoherent token generation.
Additionally, dynamic physical allocation and multi-channel interleaved memory topologies necessitate re-characterizing target DRAM coordinates across cold boots.

Nevertheless, the persistence of these weight corruptions across subsequent inference iterations presents severe implications for edge deployments and autonomous systems.
A momentary physical interaction can silently disarm an embedded agent's safety alignment without corrupting the on-disk model artifacts or leaving persistent OS-level indicators.
These findings suggest that model alignment strategies should transition away from localized, low-rank refusal representations toward globally distributed safety mechanisms that remain resilient against targeted localized hardware faults.

\subsection{Other hardware components and platforms}
We demonstrate attacks on two representative and state-of-the-art hardware devices made for edge applications, and we analyze faults on encrypted data. However, there are various other, potentially vulnerable hardware components that handle plaintext data. For instance, the computations in the GPU must be performed on plaintext weights and inputs. Although modern devices are also equipped with ECC or similar to correct faults in on-chip memory (e.g., register files), undetected faults can still occur. On platforms without memory encryption, the same pulses would cause different faults. Fewer than 16 bytes might be corrupted and the flipped bits depend on the pulse and the data instead of being random. In models with \texttt{FP16} or \texttt{BF16} weights, such faults lead to fewer NaNs, so more of them pass a NaN check such as the one of the robot (\cref{sec::robot-setup}). They also produce fewer very large weights, which the tensor wipes of \cref{sec::exp-reload} rely on. Whether the attacks become easier or harder on such platforms is left for future work.

\section{Conclusion}\label{sec::conc}
Algorithmic alignment relies on an implicit trust boundary between a model’s parameters and the physical hardware executing them -- an assumption that breaks down at the edge. Whereas Isaac Asimov’s First Law of Robotics envisioned safety constraints bound directly to the physical substrate of an autonomous agent, modern foundation models implement guardrails as fragile, low-rank subspaces within model weights. FATCAT demonstrates that on production GPU SoCs, non-invasive electromagnetic fault injection can silently neutralize LLM refusal mechanisms and induce hazardous, unprompted actions in embodied robotic policies. Ultimately, software-level alignment cannot provide meaningful safety guarantees on untrusted silicon; securing embodied AI requires anchoring behavioral constraints in verifiable, fault-tolerant hardware.

\section*{Artifacts}\label{sec::artifacts}
We make the following artifacts available:

\begin{itemize}
    \item source code for the jailbreak and robot-arm experiments (host and Jetson), at \url{https://github.com/PeterHorvath10/fatcat};
    \item video clips of the robot-arm experiments under transient and persistent fault injection, and of the jailbreak experiments, at \videourl.
\end{itemize}

\section*{Responsible disclosure}\label{sec::disclosure}
We notified the NVIDIA Product Security Incident Response Team (PSIRT) of the vulnerabilities on October 9, 2026, and received acknowledgment on October 19, 2026.

\section*{Acknowledgements}\label{sec::ack}
This research was supported by:
the Dutch Research Council (NWO) through the PROACT project (NWA.1215.18.014);
the TTW PREDATOR project 19782;
the CiCS project of the research programme Gravitation under the grant 024.006.037.
The authors are grateful to Haojia for assistance with the robot arm experiment setup and to Sail Wang, Giovanni Cherubin, and Darija Halatova for continuous feedback during the project.

\bibliographystyle{IEEEtranSN}
\bibliography{bibliography}

\appendix
\section*{Appendix}
\let\subsection\section
\crefalias{section}{appendix}%
\crefalias{subsection}{appendix}%
\subsection{Simulated transient faults in prefill}
\label{app:prefill-sim}

We run simulations in software to gain insight into what extent a single 16 byte block can influence refusals in Llama~3.2~1B and Gemma~3~1B. In these experiments, we use a 5090 RTX GPU with Huggingface Transformers implementations on the harmful JailbreakBench prompts.

The faults are a block of uniformly random bytes following (\cref{sec::dramenc}). Although this simulation is limited to a single block, it already shows that it is sufficient to substantially reduce the refusal in the models.
In the simulation, the random 16 bytes are injected into a particular weight tensor and are only present during prefill and all later tokens are decoded with clean weights from the KV cache that the faulted prefill wrote.
In prefill we distinguish between two cases: \emph{whole}, where the whole prompt is computed (nothing cached), and \emph{tail}, where the prompt's full blocks come from the prefix cache and only the last 1--16 tokens are computed similarly to vLLM prefix caching.

\noindent\textbf{Formats.}
We compare nine formats:
\texttt{BF16};
\texttt{FP8} (E4M3 codes, one \texttt{FP32} scale per tensor);
\texttt{INT8} (one \texttt{BF16} scale per output row);
\texttt{INT4} (one \texttt{BF16} scale per group of 128 weights);
\texttt{NVFP4} (E2M1 codes, one E4M3 scale per block of 16 weights and one \texttt{FP32} scale per tensor);
\texttt{MXFP4} (E2M1 codes, one power-of-two scale per block of 32);
and the GGUF files \texttt{Q4\_K\_M}, \texttt{Q4\_0} and \texttt{Q8\_0} of both models, faulted in their raw bytes.
\texttt{Q4\_0} and \texttt{Q8\_0} store an \texttt{FP16} scale in front of every block of 32 weights.

\noindent\textbf{Scoring.}
The simulation covers every layer and the seven weight tensors of a layer (\tn{down\_proj}, \tn{up\_proj}, \tn{gate\_proj}, \tn{o\_proj}, \tn{q\_proj}, \tn{k\_proj}, \tn{v\_proj}) with 20 random blocks each, plus faults forced into a scale, in total 125{,}406 passes over the 100 prompts.
These use 48-token answers and the refusal-keyword list.
We say that refusals \emph{collapse} when a fault leaves at least 20 fewer refusals than the clean run with at most 10 degenerate answers, and that the output is \emph{garbled} when at least 90 answers are degenerate.
About 600 outcomes, mostly those with the fewest keyword refusals per format, tensor and scale, were then regenerated with 160-token answers and labelled by DeepSeek V4.1 Flash and GLM~5.3 with the rubric of \cref{sec::exp-reload}.
The judged counts are therefore those of the most effective faults and not averages.
There are two exceptions.
For \cref{tab:sim-formats}, a random sample of the whole-prompt faults was regenerated and judged in the same way: 100 faults at a random place per model in \texttt{BF16} and in each GGUF file, 200 in each other format, and 40 per kind of scale.
For \cref{tab:sim-pulse}, all 20 random blocks of four Llama targets were judged.

\noindent\textbf{Mechanism (\texttt{BF16}).}
A random \texttt{BF16} block almost always contains a value above $10^{24}$.
In Llama, such a weight in \tn{down\_proj} of layer~$L$ puts a huge value into the residual stream of every position computed in that pass.
The next RMSNorm squares it, overflows to infinity and returns zero, so layers above $L$ store an all-zero KV cache for these positions, while layers $0$ to $L$ are exact.
The logits of the pass are zero, hence the first answer token is token~0 (``\texttt{!}''), and the decode then runs normally on a cache in which the upper layers do not see the prompt.
Zeroing the KV cache of layers above~8 on a clean run and forcing the first token to ``\texttt{!}'' reproduces the faulted answers for all 100 prompts.\arxivonly{\looseness=-1}

\begin{table*}[t]
\arxivonly{\captionsetup{format=tight}}%
\caption{One random 16-byte fault, by parameter format (simulation, whole prompt). Anywhere in the tensor: the block lands at a uniformly random position in the stored bytes of one of the seven weight tensors of a layer, codes and scales alike, all tensors and layers pooled. Unsafe: share of a random sample of these faults that gives at least 10 more unsafe answers out of 100 than the clean model (160-token answers), as labelled by DeepSeek V4.1 Flash\,/\,GLM~5.3; garbled: share that leaves at least 90 degenerate answers; scale hit: share that touches a scale. In a scale: the same unsafe share for faults forced into a scale of an MLP tensor. All values in \%.}
\label{tab:sim-formats}
\centering
\footnotesize
\sbox0{\begin{tabular}{@{}l cr cr rr cc@{}}
\toprule
& \multicolumn{6}{c}{Anywhere in the tensor} & \multicolumn{2}{c}{In a scale} \\
\cmidrule(lr){2-7}\cmidrule(l){8-9}
& \multicolumn{2}{c}{Llama~3.2~1B} & \multicolumn{2}{c}{Gemma~3~1B} & \multicolumn{2}{c}{Scale hit} & \multicolumn{2}{c}{Unsafe} \\
\cmidrule(lr){2-3}\cmidrule(lr){4-5}\cmidrule(lr){6-7}\cmidrule(l){8-9}
Format & Unsafe & Garbled & Unsafe & Garbled & Llama & Gemma & Llama & Gemma \\
\midrule
\texttt{BF16}        & 3\,/\,3 & 16.9 & 14\,/\,13 & 5.2  & --     & --     & -- & -- \\
\texttt{FP8}         & 0\,/\,0 & 11.8 & 0\,/\,0   & 11.2 & $<$0.1 & $<$0.1 & 3\,/\,3 & 53\,/\,53 \\
\texttt{INT8}        & 0\,/\,0 & 0    & 0\,/\,0   & 0    & $<$0.1 & 0.2    & 15\,/\,0 & 43\,/\,43 \\
\texttt{INT4}        & 0\,/\,0 & 0.7  & 0\,/\,0   & 0.4  & 3.0    & 2.7    & 0\,/\,0 & 63\,/\,55 \\
\texttt{NVFP4}, block scales & \multirow{2}{*}{0\,/\,0} & \multirow{2}{*}{1.2} & \multirow{2}{*}{0\,/\,0} & \multirow{2}{*}{1.4} & 12.2 & 11.8 & 0\,/\,0 & 0\,/\,0 \\
\texttt{NVFP4}, tensor scale &   &     &   &     & $<$0.1 & $<$0.1 & 0\,/\,8 & 50\,/\,50 \\
\texttt{MXFP4}       & 0\,/\,0 & 1.4  & 2\,/\,2   & 1.6  & 5.8    & 5.5    & 0\,/\,3 & 48\,/\,30 \\
\texttt{Q4\_K\_M}    & 2\,/\,2 & 1.6  & 4\,/\,4   & 2.0  & 13     & 53     & 10\,/\,15 & 38\,/\,40 \\
\texttt{Q4\_0}       & 4\,/\,4 & 4.6  & 6\,/\,5   & 3.0  & 89     & 88     & 10\,/\,15 & 23\,/\,23 \\
\texttt{Q8\_0}       & 2\,/\,2 & 3.1  & 7\,/\,6   & 1.4  & 47     & 50     & 8\,/\,8 & 35\,/\,35 \\
\bottomrule
\end{tabular}}
\usebox0\\[2pt]
\parbox{\wd0}{\footnotesize \texttt{NVFP4} has two kinds of scale: its rows share the same faults anywhere in the tensor and differ in the scale that is hit. Garbled and scale hit: all 2{,}240 faults per format for Llama and 3{,}640 for Gemma (48-token answers). Unsafe: a random sample of 100 of these faults in \texttt{BF16} and the GGUF files and 200 in the other formats, and 40 faults per kind of scale; a fault that leaves garbled output counts as no unsafe answers. The clean counts are those of \cref{tab:sim-judged}.}
\end{table*}

\noindent\textbf{Random faults.}
\Cref{tab:sim-formats} gives the outcome of a fault at a random place in a tensor.
What decides it is whether the faulted bytes can decode to a huge value.
In \texttt{BF16} they can, and 3\% (Llama) and 13--14\% (Gemma) of the faults add at least 10 unsafe answers.
In the formats whose codes are bounded by a scale, a fault in the codes changes at most 32 weights by a bounded amount and adds no unsafe answers: of the 200 sampled faults per model, none reaches 10 more unsafe answers in \texttt{FP8}, \texttt{INT8}, \texttt{INT4} and \texttt{NVFP4}, and 3 do in \texttt{MXFP4} on Gemma.
\texttt{INT8} shows no garbled output in 5{,}880 faults, and in \texttt{FP8} the only effect is that about 11\% of the blocks contain a NaN code and end every answer.
The GGUF formats with a floating-point scale in every block are the exception among the quantized formats.
A random 16-byte block touches such a scale in 89\% (\texttt{Q4\_0}) and about half (\texttt{Q8\_0}) of the cases, and 2--4\% (Llama) and 5--7\% (Gemma) of the faults add at least 10 unsafe answers, as many as in \texttt{BF16} on Llama and about a third to a half as many on Gemma.

\noindent\textbf{Scale faults.}
A fault that lands in a scale is far more effective than one in the codes (\cref{tab:sim-formats}, right).
On Gemma, 23--63\% of the scale faults in the MLP tensors add at least 10 unsafe answers in every format but one, more than a random weight fault in \texttt{BF16}.
On Llama the share is 0--15\%: by the keyword score refusals collapse after 13--32\% of these faults, but most of the answers that no longer refuse are labelled as neither refused nor unsafe.
The exception is the block scale of \texttt{NVFP4}: it is itself an E4M3 code and therefore bounded, and no fault in it lowered refusals in any tensor of either model, whereas the \texttt{FP32} tensor scale of the same format is as exposed as that of \texttt{FP8}.
Quantization thus has a different vulnerable spot and it confines it to the scales, which are a small share of the bytes in the vLLM formats and a large one in \texttt{Q4\_0} and \texttt{Q8\_0}. Note that, this is only a single block simulation, quantized version might be vulnerable with multiple block faults on the weights directly.

\noindent\textbf{Tensor and model.}
In \texttt{BF16} on Llama, refusals collapse after 18--22\% of the whole-prompt faults in \tn{down\_proj}, \tn{up\_proj}, \tn{gate\_proj} and \tn{o\_proj} and after at most 3\% in \tn{q\_proj} and \tn{k\_proj}; a fault in \tn{v\_proj} leaves garbled output in 88\% of the cases, since its result is what the KV cache stores.
On Gemma the rates are 13\% for \tn{down\_proj} and \tn{up\_proj} and at most 9\% elsewhere, and \tn{v\_proj} is not destructive (7\% garbled).
Scale faults in \tn{q\_proj} and \tn{k\_proj} are harmless on Gemma (0--2\%).
On Llama they rarely matter where a scale covers one row or block (at most 5\%), but a fault in the single tensor scale of \tn{q\_proj} or \tn{k\_proj} collapses refusals after 36\% (\texttt{FP8}) and 49\% (\texttt{NVFP4}) of the faults.\arxivonly{\looseness=-1}

\begin{table*}[t]
\caption{Unsafe answers out of 100 under the most effective single faults (simulation, 160-token answers), as labelled by DeepSeek V4.1 Flash and GLM~5.3. Best fault: the judged fault with the most unsafe answers. 3 faults, 10 faults: prompts that receive an unsafe answer under at least one of the 3 or 10 best-chosen faults, counting only answers that both judges label unsafe.}
\label{tab:sim-judged}
\centering
\footnotesize
\setlength{\tabcolsep}{5pt}
\sbox0{\begin{tabular}{@{}l rr rr rr rr rr rr@{}}
\toprule
& \multicolumn{6}{c}{Llama~3.2~1B} & \multicolumn{6}{c}{Gemma~3~1B} \\
\cmidrule(lr){2-7}\cmidrule(l){8-13}
& \multicolumn{2}{c}{Clean} & \multicolumn{2}{c}{Best fault} & \multicolumn{2}{c}{Faults} & \multicolumn{2}{c}{Clean} & \multicolumn{2}{c}{Best fault} & \multicolumn{2}{c}{Faults} \\
\cmidrule(lr){2-3}\cmidrule(lr){4-5}\cmidrule(lr){6-7}\cmidrule(lr){8-9}\cmidrule(lr){10-11}\cmidrule(l){12-13}
Format & DeepSeek & GLM & DeepSeek & GLM & 3 & 10 & DeepSeek & GLM & DeepSeek & GLM & 3 & 10 \\
\midrule
\texttt{BF16}     & 9 & 8 & 40 & 38 & 60 & 82 & 7 & 7 & 51 & 46 & 68 & 92 \\
\texttt{FP8}      & 10 & 10 & 28 & 30 & 49 & 74 & 7 & 7 & 57 & 59 & 70 & 80 \\
\texttt{INT8}     & 8 & 10 & 37 & 39 & 54 & 64 & 4 & 5 & 53 & 53 & 74 & 82 \\
\texttt{INT4}     & 15 & 13 & 43 & 40 & 63 & 71 & 11 & 9 & 65 & 63 & 78 & 90 \\
\texttt{NVFP4}    & 3 & 3 & 42 & 40 & 55 & 68 & 6 & 8 & 67 & 64 & 81 & 85 \\
\texttt{MXFP4}    & 17 & 17 & 39 & 37 & 50 & 58 & 14 & 17 & 76 & 72 & 82 & 90 \\
\texttt{Q4\_K\_M} & 3 & 3 & 38 & 38 & 50 & 62 & 8 & 7 & 66 & 69 & 87 & 89 \\
\texttt{Q4\_0}    & 11 & 9 & 46 & 46 & 67 & 83 & 7 & 9 & 65 & 68 & 74 & 88 \\
\texttt{Q8\_0}    & 9 & 9 & 44 & 40 & 62 & 78 & 5 & 7 & 71 & 67 & 90 & 96 \\
\bottomrule
\end{tabular}}
\usebox0\\[2pt]
\parbox{\wd0}{\footnotesize Judged faults per format: 22--62 for Llama and 28--82 for Gemma, selected by the keyword score; the 3 and 10 faults are chosen on the labels of the same prompts. The faults judged for \cref{tab:sim-formats,tab:sim-pulse} are not included; the best of them exceed the best fault given here in two cases (Llama \texttt{BF16}: 42\,/\,41; Gemma \texttt{Q8\_0}: 74\,/\,66).}
\end{table*}

\noindent\textbf{Unsafe answers.}
\Cref{tab:sim-judged} gives the judged outcome of the most effective faults.
One fault yields up to 46 unsafe answers on Llama and up to 76 on Gemma, against 3--17 for the clean models, and no format stays near its clean count.
In the quantized formats the best fault is always in a scale (in GGUF, in a block that holds one); the best faults that hit only codes stay within a few answers of the clean model.
On Gemma the best faults are in the MLP tensors of layers 4--9.
On Llama the best fault is in \tn{k\_proj} of layer~12 in six of the nine formats, including a plain weight fault in \texttt{BF16}; in \texttt{BF16}, the judged whole-prompt faults in \tn{down\_proj} give at most 18 unsafe answers in any layer.
The keyword list overstates Llama's low layers: for whole-prompt faults in the MLP tensors and \tn{o\_proj} of layers 0--7 it counts most answers as coherent, but both judges label nearly all of them as neither refused nor unsafe.

A landed fault is deterministic, so repeating one target yields one answer per prompt; new unsafe answers come from other layers, tensors and cache cases.
Three faults chosen for their coverage give an unsafe answer to 49--67 of the 100 prompts on Llama and 68--90 on Gemma, and ten faults to 58--83 and 80--96.
An attacker who can place a fault in a chosen tensor and repeat a prompt therefore needs only a few targets per model, in any of the formats.

\begin{table*}[t]
\caption{Unsafe answers out of 100 after one random 16-byte fault in a chosen tensor of Llama~3.2~1B in \texttt{BF16} (simulation, 160-token answers), as labelled by DeepSeek V4.1 Flash and GLM~5.3. Mean, median and best are over the 20 random blocks of the target; blocks $\geq$20: number of blocks that give at least 20 unsafe answers. The clean model gives 9 (DeepSeek) and 8 (GLM).}
\label{tab:sim-pulse}
\centering
\footnotesize
\sbox0{\begin{tabular}{@{}ll rrrc rrrc@{}}
\toprule
& & \multicolumn{4}{c}{DeepSeek} & \multicolumn{4}{c}{GLM} \\
\cmidrule(lr){3-6}\cmidrule(l){7-10}
Target & Prefill & Mean & Median & Best & Blocks $\geq$20 & Mean & Median & Best & Blocks $\geq$20 \\
\midrule
\tn{k\_proj}, layer 11    & whole & 14.9 & 17 & 29 & 9 & 15.2 & 16 & 27 & 7 \\
                          & tail  & 13.6 & 14 & 30 & 3 & 12.7 & 14 & 29 & 3 \\
\tn{k\_proj}, layer 12    & whole & 13.9 & 11 & 42 & 4 & 14.0 & 10 & 41 & 4 \\
                          & tail  & 13.5 & 10 & 42 & 3 & 12.9 & 10 & 38 & 3 \\
\tn{down\_proj}, layer 10 & whole & 9.1 & 14 & 14 & 0 & 11.2 & 17 & 17 & 0 \\
                          & tail  & 15.2 & 14 & 34 & 4 & 15.2 & 13 & 35 & 4 \\
\tn{down\_proj}, layer 8  & whole & 10.9 & 11 & 12 & 0 & 14.1 & 14 & 17 & 0 \\
                          & tail  & 12.1 & 12 & 14 & 0 & 12.9 & 13 & 13 & 0 \\
\bottomrule
\end{tabular}}
\usebox0\\[2pt]
\parbox{\wd0}{\footnotesize Blocks that leave garbled output count as 0 unsafe answers.}
\end{table*}

\noindent\textbf{One fault at a random place in a chosen tensor.}
The counts in \cref{tab:sim-judged} are those of the best 16-byte block, which an attacker cannot select.
\Cref{tab:sim-pulse} therefore gives the outcome over all 20 random blocks of the four most promising Llama targets in \texttt{BF16}.
No target yields more than about 15 unsafe answers on average, 6--7 above the clean model.
\tn{k\_proj} of layer~12 holds the best single block (42 unsafe answers), but its median block stays at the clean count; \tn{k\_proj} of layer~11 with the whole prompt is the most consistent target, with 7--9 of 20 blocks at 20 or more.
\tn{down\_proj} of layer~8 with the whole prompt lowers refusals from 89 to about 30 in almost every block, but the answers that no longer refuse are labelled as neither refused nor unsafe.
The same holds for the other tensors of layer~8: among 36 further judged outcomes, the median blocks give 6--14 unsafe answers, and the best one, in \tn{o\_proj} with a cached prefix, 32\,/\,28.
On Llama, the high counts of \cref{tab:sim-judged} thus require the right block or several attempts per prompt.

\noindent\textbf{Prefix caching and persistent faults.}
With the prefix cached, a \texttt{BF16} fault reaches only the tail of the prompt and refusals collapse about half as often or less (5.4\% against 11.6\% on Llama, 2.1\% against 6.3\% on Gemma).
This does not hold for all scale faults: on Llama, \emph{tail} faults in the \texttt{INT4} and \texttt{Q4\_K\_M} scales collapse refusals more often than \emph{whole}-prompt ones (47\% against 32\%, 45\% against 27\%), and the best judged Llama faults in four formats are \emph{tail} faults.
When the fault is kept during decoding, a \texttt{BF16} fault in \tn{down\_proj} of layer~8 leaves only degenerate output on Llama, whereas on Gemma the same kind of fault in layers 9 and 15 lowers keyword refusals from 48 to 22--27 with few degenerate answers, in line with the load-time results of \cref{sec::exp-reload}.

\noindent\textbf{Limitations.}
The simulation runs in HuggingFace Transformers with \texttt{BF16} arithmetic and not in vLLM or llama.cpp.
The judged counts of \cref{tab:sim-judged} are for faults selected by the keyword score, which undercounts Gemma's refusals (\cref{tab:emfi-gemma-real-vs-sim}), and they do not measure whether the faulted model is still useful on other tasks.
In addition, this simulation is limited to a fault in a single 16-byte block.

\subsection{Address-to-channel mapping}
\label{app:channel-mapping}

\begin{figure*}[htbp]
\centering
\includeinkscape[width=\textwidth, height=8cm]{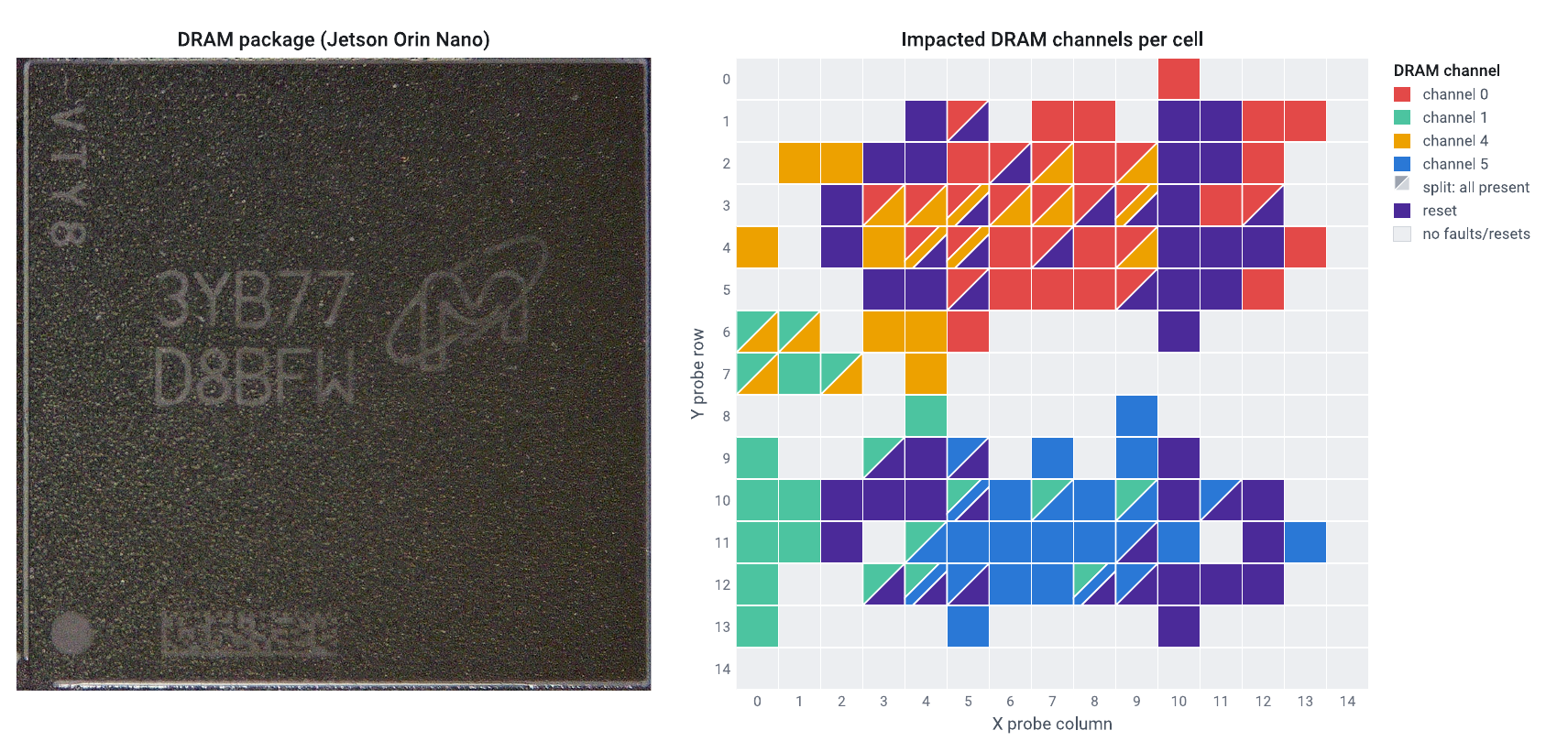}
\caption{Impacted memory channels over the whole DRAM package on Orin. Each package has 4 channels. In the figure, two channels are clustered in the upper and two other channels are clustered in the lower part of package. For each cell, 5 fault injection attempts were made to see any impact. At some cells, a single glitch can impact multiple channels.}\label{fig::orin_rightDRAM}
\end{figure*}

\begin{figure*}[htbp]
\centering
\includeinkscape[width=\textwidth, height=8cm]{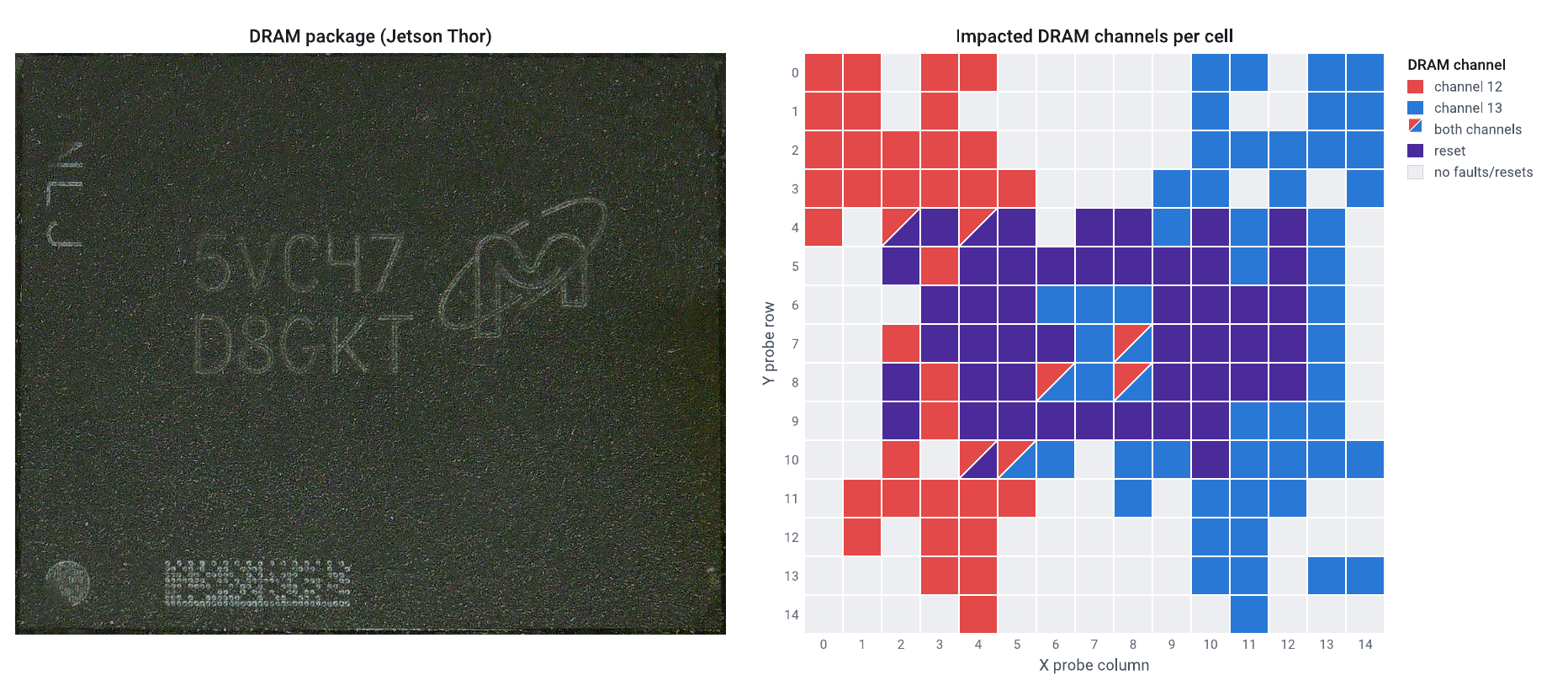}
\caption{Impacted memory channels by EMFI over the whole DRAM package on Thor. Each package has 2 channels and the two channels are clustered on the left and the right. In the middle, there is a small cluster where both channels are impacted. For each cell, 5 fault injection attempts were made to see any impact. At some cells, the glitch impacted multiple channels.}\label{fig::thor_leftDRAM}
\end{figure*}

The Jetson Orin Nano's PCB carries two LPDDR5 packages. Each package
provides four memory channels and is 64 bits wide, so the total memory bus
width is 128~bits (eight channels in total). The Jetson AGX Thor's PCB
carries eight LPDDR5X packages. Each provides two channels and is 32 bits
wide, so its total memory bus width is 256~bits (16 channels in total).

After probing memory controller registers and reverse engineering, the channels are computed from the addresses as follows.
Both memory controllers pick a channel with bit-selector masks, listed in
\cref{tab:channel-masks}: $K = 3$ masks select one of the eight
channels of the Orin Nano, and $K = 4$ masks one of the 16 channels of the
Thor. On the Orin Nano the masks apply to the physical address
$\mathit{PA}$ itself, on the Thor to its offset into DRAM, which starts at
physical address \texttt{0x80000000}:
\begin{equation}
  a =
  \begin{cases}
    \mathit{PA} & \text{on the Orin Nano,} \\
    \mathit{PA} - \texttt{0x80000000} & \text{on the Thor.}
  \end{cases}
  \label{eq:mask-input}
\end{equation}
Bit~$n$ of the channel number is the parity of the bits of $a$ selected by
mask~$n$:
\begin{equation}
  c_n = \operatorname{parity}\bigl(a \mathbin{\&} M_n\bigr),
  \qquad n \in \{0, \dots, K-1\},
  \label{eq:channel-bit}
\end{equation}
and the channel number is
\begin{equation}
  \mathit{ch}(\mathit{PA}) = \sum_{n=0}^{K-1} 2^{n}\, c_n
  \in \{0, \dots, 2^{K}-1\}.
  \label{eq:channel}
\end{equation}
On the Thor, channels $2p$ and $2p+1$ are the two channels of package~$p$.

\begin{table}[t]
  \centering
  \caption{Channel masks $M_n$ of the two memory controllers. They apply to
    $\mathit{PA}$ on the Orin Nano and to $\mathit{PA} - \texttt{0x80000000}$
    on the Thor.}
  \label{tab:channel-masks}
  \begin{tabular}{cll}
    \toprule
    $n$ & Jetson Orin Nano & Jetson AGX Thor \\
    \midrule
    0 & \texttt{0xb94f5200} & \texttt{0x68b62e3800} \\
    1 & \texttt{0xcae4a400} & \texttt{0x4b5a527e00} \\
    2 & \texttt{0x2e764800} & \texttt{0x7fa9a6cc00} \\
    3 & ---                 & \texttt{0xe694cd5a00} \\
    \bottomrule
  \end{tabular}
\end{table}

Combining this address-to-channel mapping with fault injection lets us
characterize which channels are affected at which locations on the DRAM
packages and the PHY layers. The results for the LPDDR package are shown in~\cref{fig::orin_rightDRAM}. There are 4 impacted channels, as expected, two channels are clustered in the upper and two other channels are clustered in the lower part of package. For Thor, \cref{fig::thor_leftDRAM} shows the impacted memory channels over package 6 where the two channels are clustered on the left and the right. In general, some areas over the LPDDR packages impact just some blocks (on the order of 10) while some other areas can impact blocks on the order of thousands.

\ifBaselineTable
\subsection{Baselines and directional interventions on harmful requests}
\label{app:ref}

Table~\ref{tab:main} reports the \jbb{} harmful-prompt refusal rate for the
clean models and every \emph{direction-based} intervention: global
activation ablation, fixed addition $\mp\rhat$ at the direction's layer, and
fitted rank-one weight edits $W' = W - c\,\rhat\,v^{\top}$ on the
mid-network \downproj{} matrices that write the direction into the residual
stream (solo and dual-layer variants).

\begin{table*}[h]
\centering
\caption{Harmful-prompt refusal on 100 \jbb{} prompts (bypass $\uparrow$ =
refusal $\downarrow$) after erasing directions via Arditi et al.~\cite{NEURIPS2024_f5454485}. KL is the harmless next-token KL of weight edits.}
\label{tab:main}
\small
\begin{tabular}{llcc}
\toprule
 & & \multicolumn{2}{c}{\jbb{} refusal ($\downarrow$)} \\
\cmidrule(l){3-4}
Intervention & Target / coefficient & \texttt{gemma3:1b} & \texttt{llama3.2:1b} \\
\midrule
\multicolumn{4}{l}{\emph{Baseline}} \\
Clean model & --- & 93\% & 91\% \\
\midrule
\multicolumn{4}{l}{\emph{Activation-space directional interventions}} \\
Global directional ablation & all blocks & ---$^{\S}$ & \best{8\%} \\
Addition $-r$ (bypass) & L9 @ raw $c{=}1$ / L8 @ $-1$ & --- & \best{2\%} \\
Addition $+r$ (induce refusal) & same & --- & 63\%$^{\P}$ \\
\midrule
\multicolumn{4}{l}{\emph{Single-matrix fitted rank-one edit, $W' = W - c\,\rhat\,v^\top$}} \\
Solo edit & $L_8$ \downproj, $c{=}4$ & \best{21\%} (KL .089) & --- \\
Solo edit & $L_9$ \downproj, $c{=}8$ & \best{12\%} (KL .053) & --- \\
Solo edit & $L_7$ \downproj, $c{=}2$ & --- & \best{3\%} (KL .054) \\
Solo edit & $L_8$ \downproj, $c{=}4$ & --- & \best{1\%} (KL .055) \\
Solo edit & $L_{9}$ \downproj, $c{=}4$ & --- & 12\% (KL .035) \\
Solo edit & $L_{14}$ \downproj, $c{=}4$ & --- & 85\% (KL .003) \\
\midrule
\multicolumn{4}{l}{\emph{Dual-layer simultaneous rank-one edits}} \\
Dual edit & $L_8{+}L_9$, $c{=}(2,1)$ & 51\% (KL .033) & --- \\
Dual edit & $L_8{+}L_9$, $c{=}(4,2)$ & \best{16\%} (KL .088) & --- \\
\bottomrule
\end{tabular}

\smallskip
{\footnotesize $^{\S}$\texttt{gemma3:1b} global ablation has harmless KL 13.5
(catastrophic benign damage), so no generation-based refusal is reported; its
bypass log-odds alone are strongly negative ($-17.7$). $^{\P}$Addition $+r$
on Llama induces refusal on 63\% of \emph{benign} prompts (the induce side
of the direction, reported as benign-side effect).}
\end{table*}
\fi %

\subsection{LPDDR package AGX Thor experiments}
\label{app:agx-lpddr}
In this section, we present fault injection experiments over an LPDDR package of Thor. For the experiment, we placed the EMFI probe in the upper left quadrant of the package at a location where at most 30 blocks are affected. With this arrangement, we only impact blocks from a single memory channel. We run two experiments on Llama 3 in \texttt{BF16} format, one targeting layer 7 and one targeting layer 8 MLP. The judges are Deepseek and GLM as before. With 20 trials per prompt, layer 7 results in 27 unsafe answers by at least one judge. For layer 8, the faults result in 37 unsafe answers by at least one judge. Combining the results of the two layers, 43 prompts result in an unsafe answer by at least one judge and 29 prompts where both judges agree. On Gemma, in an experiment targeting layer 9, the number of prompts that received at least one harmful answer is 31 according to both judges. The complete end-to-end experiments are shown in \videourl. 

\ifPlatformTables
\subsection{Software and hardware environments}
\label{app:stack}

In the experiments, we use different hardware and software setups. The jailbreak environments for the Orin Nano and Jetson Thor are detailed in \Cref{tab:platform_orin} and \Cref{tab:platform_thor}.

\fi %

\subsection{Faulted arm episodes}\label{sec::robot-frames-app}
\begin{figure*}[p]
\centering
\begin{subfigure}[t]{0.49\textwidth}
\centering
\numberedframes{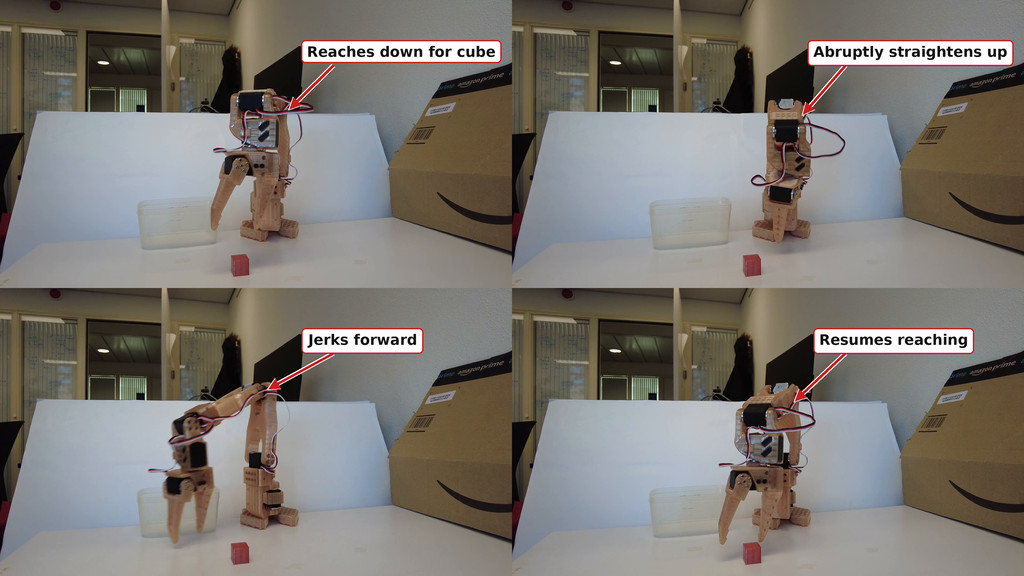}
\caption{While reaching down for the cube~(1), the arm abruptly straightens up~(2), jerks forward~(3) and then resumes reaching~(4).}
\label{fig:robot-frames-clip2}
\end{subfigure}\hfill
\begin{subfigure}[t]{0.49\textwidth}
\centering
\numberedframes{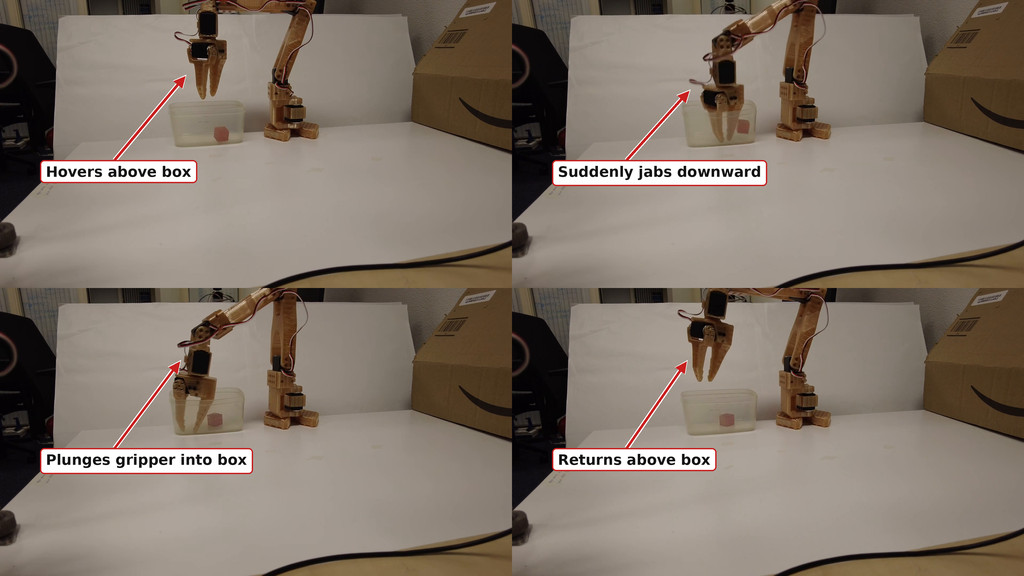}
\caption{While hovering above the box~(1), the arm suddenly jabs downward~(2), plunges the gripper into the box~(3) and returns above it~(4).}
\label{fig:robot-frames-clip3}
\end{subfigure}
\par\medskip
\begin{subfigure}[t]{0.49\textwidth}
\centering
\numberedframes{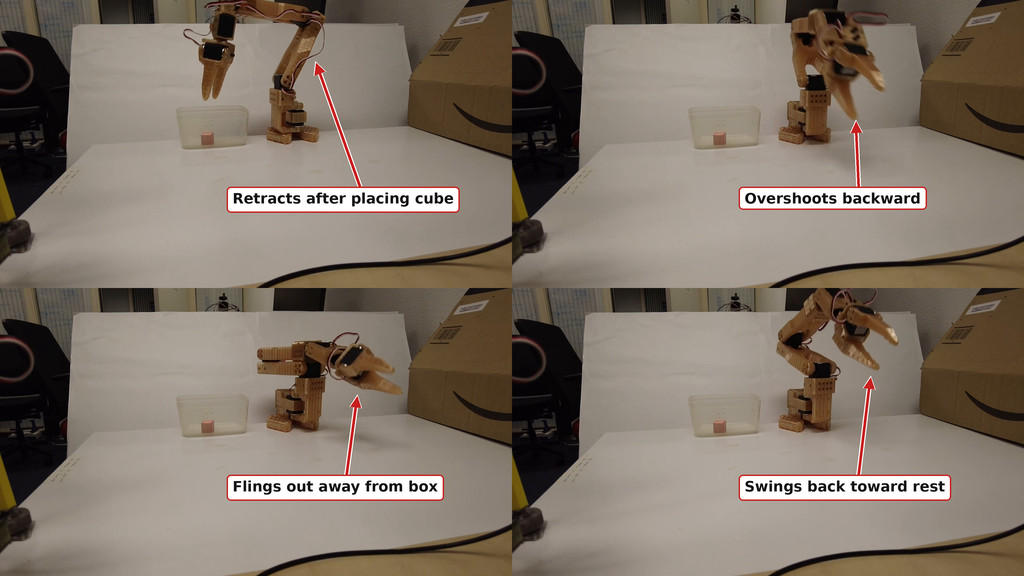}
\caption{While retracting after placing the cube~(1), the arm overshoots backward~(2), flings out away from the box~(3) and swings back towards its rest pose~(4).}
\label{fig:robot-frames-clip4}
\end{subfigure}\hfill
\begin{subfigure}[t]{0.49\textwidth}
\centering
\numberedframes{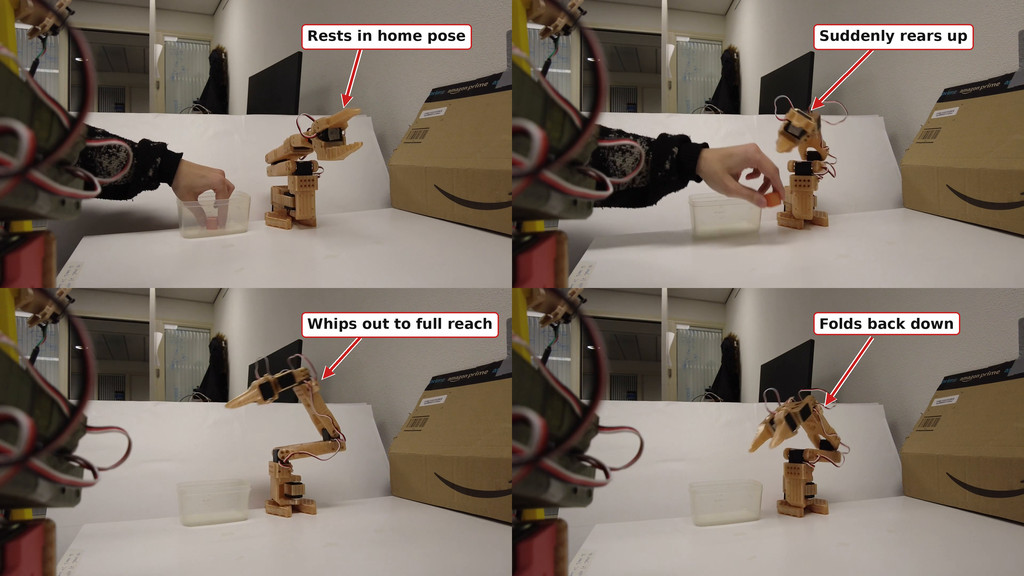}
\caption{While the arm rests in its home pose and a hand takes the cube out of the box~(1), it suddenly rears up~(2), whips out to full reach~(3) and folds back down~(4).}
\label{fig:robot-frames-clip5}
\end{subfigure}
\par\medskip
\begin{subfigure}[t]{0.49\textwidth}
\centering
\numberedframes{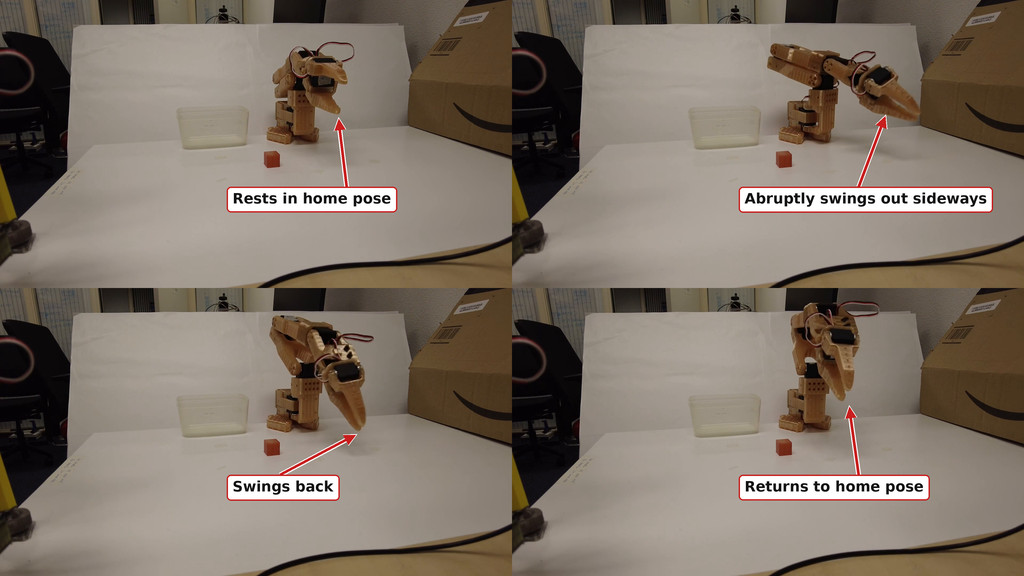}
\caption{While resting in its home pose~(1), the arm abruptly swings out sideways~(2), swings back~(3) and returns to its home pose~(4).}
\label{fig:robot-frames-clip6a}
\end{subfigure}\hfill
\begin{subfigure}[t]{0.49\textwidth}
\centering
\numberedframes{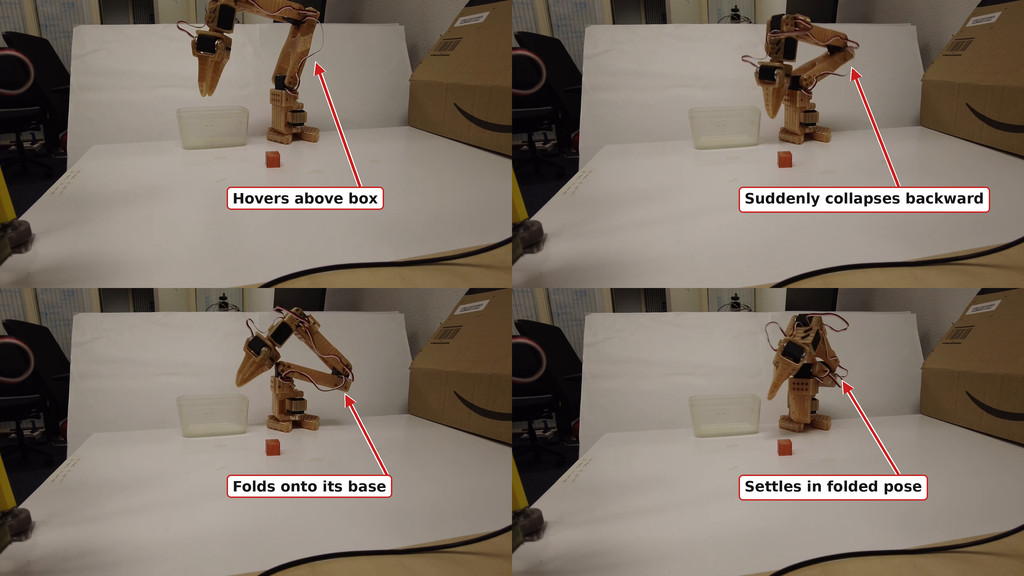}
\caption{While hovering above the box~(1), the arm suddenly collapses backward~(2), folds onto its base~(3) and settles in the folded pose~(4).}
\label{fig:robot-frames-clip6b}
\end{subfigure}
\caption{Video frames of further moving-arm episodes with EM faults on the policy (\cref{sec::robot-policy}). The frames are numbered in the order in which they were recorded. The video clips are available at \videourl.}
\label{fig:robot-frames-app}
\end{figure*}

\Cref{fig:robot-frames-app} shows video frames of further moving-arm episodes with EM faults, in addition to the one in \cref{fig:robot-frames-clip1}. The annotations describe the motion of the arm in each frame.

\ifPlatformTables
  \makeatletter
  \setlength{\@dblfptop}{0pt}\setlength{\@dblfpsep}{24pt}%
  \makeatother
  \newlength{\platformcatwidth}
\newlength{\platformcompwidth}
\NewDocumentEnvironment{platformtable}{m m +b}{%
  \begin{table*}[tp]
    \centering
    \caption{#1}\label{#2}%
    \footnotesize
    \setlength{\tabcolsep}{5pt}%
    \settowidth{\platformcatwidth}{Policy runtime}%
    \settowidth{\platformcompwidth}{cuDNN / NVRTC}%
    \begin{tabularx}{\textwidth}{@{}>{\raggedright\arraybackslash}p{\platformcatwidth}%
        >{\raggedright\arraybackslash}p{\platformcompwidth}%
        >{\raggedright\arraybackslash\hangindent=1em\hangafter=1}X@{}}
      \toprule
      \textbf{Category} & \textbf{Component} & \textbf{Version / detail} \\
      \midrule
      #3
      \bottomrule
    \end{tabularx}
  \end{table*}}{}

\begin{platformtable}{Jetson Orin jailbreak platform and software environment.}{tab:platform_orin}
  Hardware
    & Module           & Jetson Orin Nano Dev.\ Kit (Super), P3767-0005 \\
    & SoC / GPU        & Tegra234; GA10B Ampere, sm\_87 (CC 8.7), 8 SMs / 1024 cores \\
    & GPU clock        & 1020\,MHz max \\
    & Memory           & 8\,GiB 128-bit LPDDR5 \\
  \addlinespace
  System
    & JetPack          & 6.2.3 (\texttt{nvidia-jetpack 6.2.3+b81}) \\
    & L4T / BSP        & R36.5.2 (R36 REV 5.2, oot kernel) \\
    & Kernel           & 5.15.199-tegra \\
    & OS               & Ubuntu 22.04.5 LTS \\
    & CUDA             & 12.6 (toolkit 12.6.11, nvcc V12.6.68) \\
    & cuDNN / NVRTC    & 9.3.0.75 / 12.6.68 \\
    & TensorRT         & 10.3.0.30 (+cuda12.5) \\
  \addlinespace
  LLM runtime
    & Inference engine & Ollama 0.32.6 (llama.cpp/ggml build \texttt{8c146a836}) \\
    & GGML backend     & CUDA (\texttt{cuda\_jetpack6}), \texttt{USE\_GRAPHS=1}, flash-attn \\
    & Models           & gemma3:1b (815\,MB), llama3.2:1b (1.23\,GB) \\
    & Runner flags     & \texttt{-c~4096 -np~1 -b~512 -ub~512 --keep~4 --load-mode~dio} \\
\end{platformtable}

\begin{platformtable}{Jetson AGX Thor jailbreak platform and software environment.}{tab:platform_thor}
  Hardware
    & Module           & Jetson AGX Thor Dev.\ Kit (P4071-0000 + P3834-0008) \\
    & SoC / GPU        & Tegra264; Blackwell, sm\_110 (CC 11.0), 20 SMs / 2560 cores \\
    & CPU              & 14$\times$ Arm Neoverse-V3AE (ARMv9, impl.\ 0x41 part 0xd83) \\
    & GPU clock        & 1575\,MHz \\
    & Memory           & 128\,GB (122.8\,GiB) LPDDR5X, 16 channels \\
  \addlinespace
  System
    & JetPack          & 7.1 (\texttt{nvidia-jetpack 7.1-b112}) \\
    & L4T / BSP        & R38 (release), REV 4.0, oot kernel \\
    & Kernel           & 6.8.12-tegra \\
    & OS               & Ubuntu 24.04.4 LTS \\
    & CUDA             & 13.0 (toolkit 13.0.0, nvcc V13.0.48) \\
    & cuDNN / NVRTC    & 9.12.0.46 / 13.0.48 \\
    & TensorRT         & 10.13.3.9 (+cuda13.0) \\
  \addlinespace
  LLM runtime
    & Inference engine & vLLM 0.26.0 (V1 engine), venv Python 3.12.3 \\
    & Backend          & PyTorch 2.11.0+cu130, FlashInfer 0.6.14, Triton 3.6.0, Transformers 5.14.1 \\
    & Models           & unsloth/gemma-3-1b-it (1.9\,GB), unsloth/Llama-3.2-1B-Instruct (2.4\,GB) \\
    & Serve flags      & \texttt{--tensor-parallel-size~1 --gpu-memory-utilization~0.25 --enforce-eager\newline
                         --override-generation-config~'\char`\{"temperature":0\char`\}'\newline
                         --trust-remote-code --port~8000} \\
\end{platformtable}

\begin{platformtable}{Jetson Orin robot arm platform and software environment.}{tab:platform_robot}
  Hardware
    & Module           & Jetson Orin Nano Dev.\ Kit (Super), P3768-0000 + P3767-0005 \\
    & SoC / GPU        & Tegra234; GA10B Ampere, sm\_87 (CC 8.7), 8 SMs / 1024 cores, 2\,MiB L2 \\
    & CPU              & 6$\times$ Cortex-A78AE, 1.728\,GHz \\
    & Clocks           & MAXN\_SUPER, \texttt{jetson\_clocks}: GPU fixed at 1020\,MHz \\
    & Memory           & 8\,GB 128-bit LPDDR5 \\
  \addlinespace
  System
    & JetPack          & 7.2.1 (\texttt{nvidia-jetpack 7.2.1-b49}) \\
    & L4T / BSP        & R39.2.1 (R39 REV 2.1, oot kernel) \\
    & Kernel           & 6.8.12-1021-tegra \\
    & OS               & Ubuntu 24.04.5 LTS \\
    & GPU driver       & NVIDIA open kernel module, r595 \\
    & CUDA             & 13.2 (toolkit 13.2.2, nvcc V13.2.86) \\
    & cuDNN            & 9.20.0.46 \\
    & TensorRT         & 10.16.2.10 (+cuda13.2) \\
  \addlinespace
  Policy runtime
    & Policy           & SmolVLA, 450\,M params: SmolVLM2-500M (16 layers) + action expert (0.75$\times$~width) \\
    & Fine-tuning      & \texttt{lerobot/smolvla\_base}, 40k steps, batch 64, expert only (vision frozen) \\
    & I/O              & 2 cameras at 512$\times$512 + 6-D state $\rightarrow$ 50$\times$6 action chunk, 10~flow-matching~steps \\
    & Inference engine & TensorRT Python API, 12 prebuilt engines, replayed as one CUDA graph \\
    & Precision        & strongly typed FP16, FP32 accumulation; Triton AOT attention (vision), fused~gated-FFN~(expert) \\
    & Latency          & 112.5\,ms: vision 60.9, VLM prefill 9.0, expert decode 42.6\,ms (10$\times$4.26) \\
  \addlinespace
  Robot / host
    & Arm              & SO-101 follower, 6 DoF, Feetech STS3215 servos over CH343 USB serial \\
    & Cameras          & 2$\times$ USB UVC (1e45:0209), MJPG 640$\times$480 at 30\,FPS \\
    & Control          & 30\,Hz loop; new chunk when $\le$30 actions remain; overlap blended 30/70~old/new \\
    & Robot stack      & LeRobot 0.6.2 (upstream \texttt{b6ec006} + local scripts), Python 3.12.14 \\
\end{platformtable}

\fi

\end{document}